\documentclass{emulateapj}
\usepackage{epsfig}
\usepackage[figuresright]{rotating}
\usepackage{natbib}
\usepackage{xcolor}
\usepackage[colorlinks, citecolor=blue]{hyperref}
\usepackage{longtable}
\usepackage{url}
\usepackage{lineno}
\usepackage{amsmath}

\shorttitle{Thermal and nonthermal emission of GRB 211211A}
\shortauthors{Chang et al.}
\slugcomment{}

\begin{document}

\title{Thermal and Nonthermal Emission from a Peculiar Long-duration GRB 211211A}
\author{Xue-Zhao Chang\altaffilmark{1}, Hou-Jun L\"{u}\altaffilmark{1}, Xing Yang\altaffilmark{1}, Jia-Ming
Chen\altaffilmark{2}, and En-Wei Liang\altaffilmark{1}} \altaffiltext{1}{Guangxi Key Laboratory for
Relativistic Astrophysics, School of Physical Science and Technology, Guangxi University, Nanning
530004, China; lhj@gxu.edu.edu} \altaffiltext{2}{School of Physics and Astronomy, Yunnan
University, Kunming 650500, China}

\begin{abstract}	
Long-duration GRB 211211A that lacks a supernova emission even down to very stringent limits at
such a low redshift $z=0.076$ and is associated with kilonova emission, suggests that its physical
origin is from a binary compact star merger. By reanalyzing its data observed with the Gamma-Ray
Burst Monitor on board the Fermi mission, we find that both time-integrated and time-resolved
spectra can be fitted well by using a 2SBPL plus blackbody (2SBPL+BB) model in the prompt emission.
The bulk Lorentz factors ($\Gamma_{\rm ph}$) of the outflow can be inferred by invoking the
observed thermal emission at the photosphere radius within a pure fireball model, and we find out
that the temporal evolution of $\Gamma_{\rm ph}$ seems to be tracking with the light curve. The
derived values of $\Gamma_{\rm ph}$ are also consistent with the $\Gamma_{\rm ph}$-$L_{\gamma, \rm
iso}$/$E_{\gamma, \rm iso}$ correlations that had been found in other bursts. Moreover, we also
calculate the magnetization factor $\sigma_{0}$ in the central engine and $\sigma_{\rm ph}$ at the
photosphere radius within the framework of a hybrid jet model, and find that the values of both
$1+\sigma_{\rm 0}$ and $1+\sigma_{\rm ph}$ are larger than 1 for different time slices. It suggests
that at least the Poynting-flux component is indeed existent in the outflow. If this is the case,
one possible physical interpretation of thermal and nonthermal emissions in GRB 211211A is from the
contributions of both $\nu\bar{\nu}$ annihilation and the Blandford-Znajek mechanisms in the
relativistic jet when a stellar mass black hole resides in the central engine.
\end{abstract}


\section{Introduction}
The field that studies of gamma-ray bursts (GRBs) has made a great progress in both observations
and theories \citep[see][for a review]{2015PhR...561....1K}. In general, GRBs are thought to be
from a catastrophic event (such as massive star core collapse or the merger of two compact stars)
to release its gravitational energy that is converted in the form of thermal energy, and a central
engine (black hole or neutron star) is formed after the catastrophic event
\citep{Eichler1989,Usov1992,Thompson1994,Dai1998a,Dai1998b,1999ApJ...518..356P,Narayan2001,
Zhang2001,2009ApJ...700.1970L,Metzger2011,Bucciantini2012,Lv2014,Berger2014,Lv2015}. The fireball
model is the most popular one to interpret both $\gamma$-ray emission and broadband afterglow
emission of GRB phenomena
\citep{1997ApJ...476..232M,1998ApJ...497L..17S,2002ARA&A..40..137M,2004IJMPA..19.2385Z,Zhang2016}.
The observed prompt emission can be explained by the photosphere with a quasi-thermal spectrum
\citep{1986ApJ...308L..43P,1986ApJ...308L..47G}, or internal shocks with a nonthermal
electromagnetic radiation \citep{1994ApJ...430L..93R}, or dissipation of the magnetic energy with a
nonthermal spectrum at a large radius \citep{2011ApJ...726...90Z}.

Within the above scenario, the photons that are produced in the thermalized outflow can escape near
the photospheric radius when the optical depth is  close to 1, and a purely quasi-thermal component
should be observed in the prompt emission of a GRB, such as GRB 090902B
\citep{2009ApJ...706L.138A,2010ApJ...709L.172R,2011ApJ...730..141Z}. On the contrary, if the
outflow is Poynting-flux dominated, it should produce a purely nonthermal emission, such as GRB
080916C \citep{2009ApJ...700L..65Z}. In general, the dissipated radius of the photospheric and
internal shock cannot be distinguished very well, and the observed spectrum of GRB prompt emission
should be the superposition of thermal and nonthermal components
\citep{2005ApJ...625L..95R,2015ApJ...801...2}. Such evidence of thermal and nonthermal emission is
already found in several solid cases, e.g., GRB 100724B \citep{2011ApJ...727L..33G}; GRB 110721A
\citep{2012ApJ...757L..31A}; GRB 120323A \citep{2013ApJ...770...32G}; GBR 160625B
\citep{2017ApJ...849...71L}; GRB 081221 \citep{2018ApJ...866...13H}.

Most recently, a peculiar and nearby long-duration GRB 211211A that triggered the Fermi Gamma-Ray
Burst Monitor (GBM; \citealt{2021GCN.31210....1M}), Swift Burst Alert Telescope (BAT;
\citealt{2021GCN.31202....1D}), as well as Insight-HXMT \citep{2021GCN.31236....1Z}, is very
excited for attention with redshift $z$=0.076. The light curve of prompt emission is composed of an
initial hard-main emission (with a duration $\sim 13$ s) followed by a series of soft gamma-ray
extended emission (EE) with a duration $\sim 55$ s, and the structure of the light curve is similar
to the particularly interesting case GRB 060614\citep{yang2022,xiao2022} and GRB 211227A
\citep{2022ApJ...931L..23L}. More interestingly, no associated supernova signature is detected for
GRB 211211A, even down to very stringent limits at such a low redshift, but associated with
kilonova is observed by several optical telescopes \citep{Rastinejad2022}. That observed evidence
suggests that GRB 211211A is originated from a binary compact star merger
{\citep{Rastinejad2022,yang2022,xiao2022,Gompertz2022}}. However, how to produce such long-duration
emission within the compact star merger scenario remains an open question. \citet{Gao2022} proposed
that the black hole central engine surrounded by a strong magnetic flux can well interpret the
behavior of long-duration emission of GRB 211211A. \citet{Gompertz2022} found that the spectrum can
be fitted well with a double smoothly broken power-law model SBPL, which is interpreted as
synchrotron emission, including both characteristic synchrotron frequency ($\nu_m$) and the cooling
frequency ($\nu_c$). So that identifying the composition of the jet in such a binary system will
play an important role in understanding the physical process and mechanism
\citep{2011ApJ...730..141Z,2015PhR...561....1K,2018pgrb.book.....Z}.

In this paper, by analyzing the data observed with the GBM on board the Fermi mission, we find that
a blackbody emission with a nonthermal component 2SBPL function can be fitted well to the spectra
in the prompt emission of GRB 211211A, especially in the initial hard-main emission phase. It means
that the thermal component should be indeed existent, and it is different from previous studies
that claimed the single nonthermal component is dominated in the prompt emission phase
\citep{yang2022,xiao2022}. The data analysis and spectral fitting are presented in \S 2. In \S 3,
we derive the Lorentz factor of the jet, its photospheric radius, magnetization $\sigma_0$, and the
dimensionless entropy $\eta$ based on the observed thermal and nonthermal emissions of GRB 211211A.
The conclusions are drawn in \S 4 with some additional discussion. Throughout the paper, a
concordant cosmology with parameters $H_0=70~\rm km~s^{-1}~Mpc^{-1}$, $\Omega_M=0.30$, and
$\Omega_{\Lambda}=0.70$ is adopted.

\section{Fermi/GBM data analysis}
\subsection{Light curve and spectral fits}
GRB 211211A triggered the Fermi/GBM at 13:09:59.651 UT on 2021 December 11
\citep{2021GCN.31210....1M}. This GRB was also detected by Swift/BAT \citep{2021GCN.31202....1D}
and Insight-HXMT \citep{2021GCN.31236....1Z}. We downloaded the corresponding time-tagged-event
(TTE) data of GRB 211211A from the public science support center at the official Fermi
website\footnote{https://heasarc.gsfc.nasa.gov/FTP/fermi/data/gbm/daily/}. The GBM has 12 sodium
iodide (NaI) detectors covering an energy range from 8 keV to 1 MeV, and two bismuth germanate
(BGO) scintillation detectors sensitive to higher energies between 200 keV and 40 MeV
\citep{2009ApJ...702..791M}. We select the brightest NaI and BGO detectors for the analyses, namely
n2, na and b0. For more details of data reduction of the light curve, please refer to
\citet{2017ApJ...849...71L} and \citet{2018NatAs...2...69Z}.

\begin{figure}
\centering
 \includegraphics [angle=0,scale=0.5] {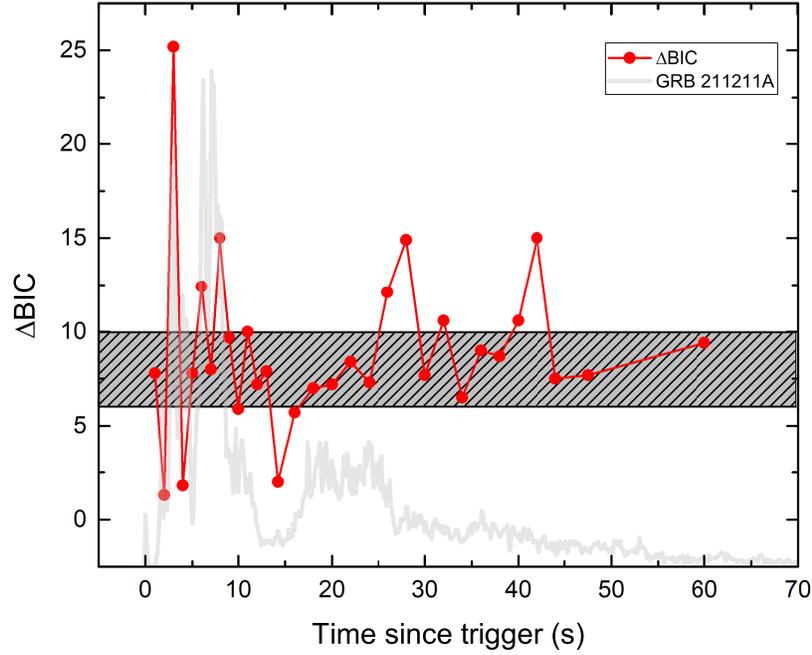}
 \caption{$\Delta BIC$ (solid red circles) between 2SBPL and 2SBPL+BB models in the time-resolved spectra
 of GRB 211211A. The gray line is the light curve of prompt emission, and the shaded area indicates the
 range of [6-10].}
 \label{fig:LCBIC}
\end{figure}

We extract the light curves with a 128 ms time bin (Figure \ref{fig:LCBIC}) by running {\em gtbin}.
The light curve shows a complex structure with a total duration of about $T_{90}\sim$43 s, an
initially main emission (with a duration $\sim 13$ s) followed by a series of soft gamma-ray
extended emission with a duration $\sim 55$ s \citep[also see][]{yang2022}.

Both time-integrated and time-resolved spectra of this source are extracted from the TTE data. The
background spectrum from the GBM data is extracted from the continuous spectroscopy (CSPEC) format
data with two time intervals before and after the prompt emission phase and are modeled with a
polynomial function. We perform the spectral fit with the multimission maximum likelihood framework
package (\citealt{2017ifs..confE.130V}), which adopts the Markov Chain Monte Carlo (MCMC) technique
to perform time-resolved spectral fitting. Also, we evaluate the goodness of our fits with the
maximum likelihood based statistics, the so-called PGSTAT. \citet{yang2022} invoked a cutoff
power-law model to do the spectral fitting in both the time-averaged and time-resolved spectra of
GRB 211211A {\citep[also see][]{Gompertz2022}}. In our analysis, several spectral models can be
selected to test the spectral fitting of the burst, such as cutoff power-law (CPL), Band function
(Band), a smoothly broken power-law model (SBPL), a double smoothly broken power-law model (2SBPL)
and blackbody (BB), as well as combinations of any two models. The Band function
\citep{1993ApJ...413..281B} and blackbody function, and CPL models are written as follows:
 \begin{eqnarray}
N_{\textrm{Band}}(E)=A\left\{\begin{array}{clcc}
(\frac{E}{100~\mathrm{keV}})^{\alpha }\mathrm{exp}\left[-\frac{(\alpha+2)E}{E_{p}} \right ], E<E_{c}, \\
(\frac{E}{100~\mathrm{keV}})^{\beta }\mathrm{exp}(\beta -\alpha
)(\frac{E_{c}}{100~\mathrm{keV}})^{\alpha-\beta }, E\geq E_{c}
\end{array}\right.
\end{eqnarray}
where $A$ is the normalization of the spectrum, $\alpha$ and $\beta$ are the low and high-energy
photon spectral indices, respectively,and $E_{p}=(2+\alpha)E_{c}$ is the peak energy;
\begin{equation}
N_{\textrm{BB}}(E)=A(t) \frac{E^{2}}{\exp [E / k T]-1}
\end{equation}
where $k$ and $T$ are Boltzmann constant and temperature, respectively;
\begin{eqnarray}
N_{\rm CPL}(E) = A\cdot E^{-\alpha}\rm exp(-\frac{E}{E_{\rm p}}).
\end{eqnarray}
The SBPL function \citep{2018A&A...613A..16R} function is defined as follows:
\begin{eqnarray}
N_{\mathrm{SBPL}}(E)=A\left(\frac{E}{100~\mathrm{keV}}\right)^{b} 10^{\left(a-a_{\mathrm{piv}}\right)}
\end{eqnarray}
where $a=m \Lambda \ln \left(\frac{e^{q}+e^{-q}}{2}\right)$, $a_{\mathrm{piv}}=m \Lambda \ln
\left(\frac{e^{q_{\mathrm{piv}}}+e^{-q_{\mathrm{piv}}}}{2}\right)$, $q=\frac{\log \left(E /
E_{b}\right)}{\Lambda}$,
 $q_{\mathrm{piv}}=\frac{\log \left(100~\mathrm{keV} / E_{b}\right)}{\Lambda}$, $m=\frac{\beta-\alpha}{2}$,
 $b=\frac{\alpha+\beta}{2}$. Here, $\alpha$, $\beta$ and $E_b$ are the lower power-law index, upper power-law 
 index,
 and a break energy, respectively; $\Lambda$ is the break scale, which is fixed at
0.3.

The 2SBPL function \citep{2006ApJS..166..298K} function is defined as follows:
\begin{eqnarray}
\begin{aligned}
N_{\mathrm{2SBPL}}=A E_{\mathrm{b}}^{\alpha_{1}}[f_1(E)+f2(E)]^{-\frac{1}{n_{2}}}
\end{aligned}
\end{eqnarray}
where$f_1(E)=[(\frac{E}{E_{\mathrm{b}}})^{-\alpha_{1}
n_{1}}+(\frac{E}{E_{\mathrm{b}}})^{-\alpha_{2} n_{2}}]^{n_2/n_1}$,
$f_2(E)=(\frac{E}{E_{\mathrm{j}}})^{-\beta
n_2}[(\frac{E_{\mathrm{j}}}{E_{\mathrm{b}}})^{-\alpha_{1}
n_{1}}+(\frac{E_{\mathrm{j}}}{E_{\mathrm{b}}})^{-\alpha_{2} n_{2}}]^{n_2/n_1}$, and
$E_{\mathrm{j}}=E_{\text {pk }}
\cdot\left(-\frac{\alpha_{2}+2}{\beta+2}\right)^{\frac{1}{(\left.\beta-\alpha_2\right) n_{2}}}$;
$\alpha_{1}$ and $\alpha_{2}$ are the photon index below and above the break energy, respectively.
$E_{b}$ and $E_{pk}$ are the break energy and peak energy, respectively; $\beta$ is the high-energy
photon index above the peak energy; $n_{1}$ (for the break) and $n_{2}$ (for the peak) are the
smoothness parameters. In this work, we fix the $n_{1}=5.38$ and $n_{2}=2.69$, respectively
\citep{2018A&A...613A..16R}.

\begin{figure}
\centering
 \includegraphics [angle=0,scale=0.45] {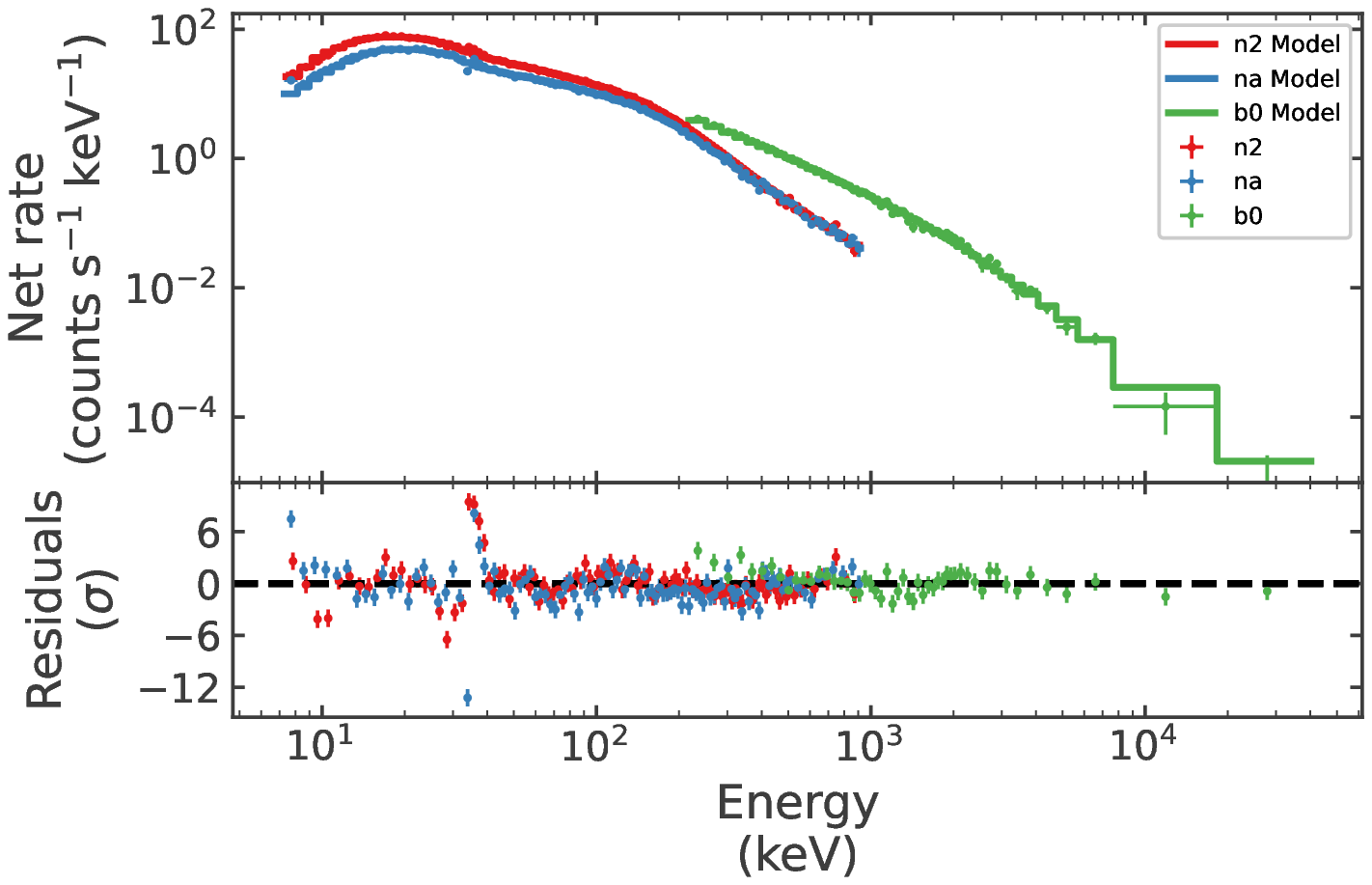}
\includegraphics [angle=0,scale=0.2] {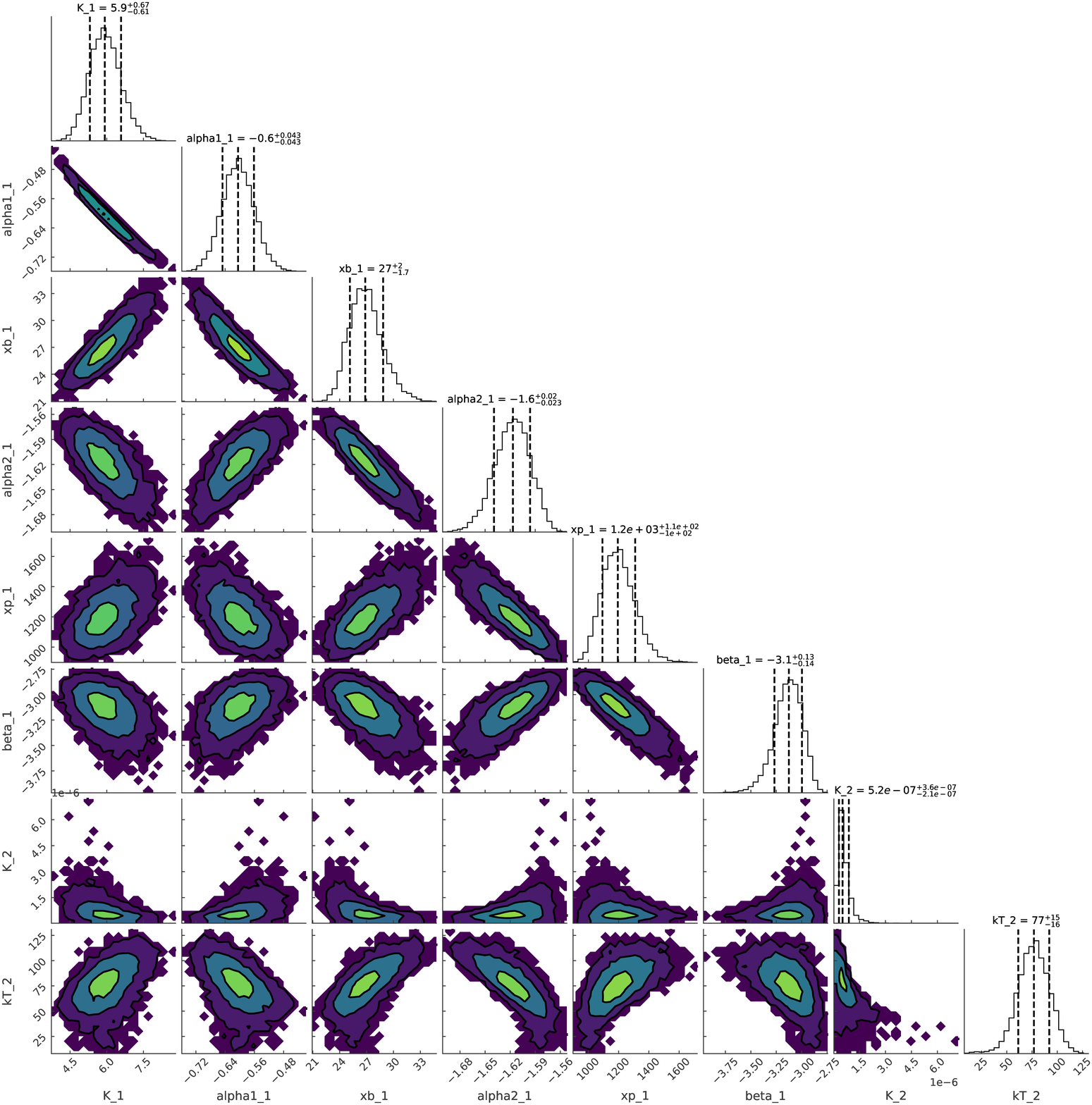}
 \caption{Time-integrated spectrum measured from $T_0+0.5$ to $T_0+70$ s is fitted by 2SBPL+BB model.
 Left: observed and modeled photon count spectra. Right: the parameter constraints of the spectral fit.}
 \label{fig:Spec}
\end{figure}

In order to test which model is the best fit with the data, we compare the goodness of the fits by
invoking the Bayesian information criteria (BIC)\footnote{BIC is a criterion to evaluate the best
model fit among a finite set of models, and the lowest BIC of a model is preferred
{\citep{Neath2012}}. The definition of BIC can be written as: $BIC=\rm -2ln L+k\cdot ln(n)$, where
$k$ is the number of model parameters, $n$ is the number of data points, and $L$ is the maximum
value of the likelihood function of the estimated model. (1) if $0<\Delta BIC<2$, the evidence
against the model with higher BIC is not worth more than a bare mention; (2) if $2<\Delta BIC<6$,
the evidence against the model with higher BIC is positive; (3) if $6<\Delta BIC<10$, the evidence
against the model with higher BIC is strong; (4) if $10<\Delta BIC$, the evidence against the model
with higher BIC is very strong.}. For the specific definition of the goodness of data fitting by
the empirical model, please refer to \citet{Li2019}. We find that the 2SBPL+BB model is the best
one to adequately describe the observed data, and it means that the thermal emission component in
GRB 211211A is a significant presence (see Table 1). The spectral fitting result of the
time-integrated spectra is shown in Figure \ref{fig:Spec}. Moreover, we find that the photon
indices of the time-integrated spectra are $\alpha_{1}=-0.6$ and $\alpha_{2}=-1.62$, respectively.
These values are consistent with the results in \citet{Gompertz2022}, and it suggests that the
nonthermal component may be originated from the synchrotron emission in the fast regime.

We also extract time-resolved spectral analyses of GRB 211211A between $T_{0}$ + 0.5 and $T_{0}$ +
70. We divide the time interval into 31 slices, and fit those slices by invoking 2SBPL and 2SBPL+BB
models. The fitting results are shown in Table 2. Figure \ref{fig:LCBIC} shows the evolution of
$\Delta$BIC, which is defined as $\rm \Delta BIC=BIC_{2SBPL}-BIC_{2SBPL+BB}$. We find that the
$\Delta$BIC of all time slices is larger than zero, especially, the $\Delta$BIC at the peak of
light curve in both the main emission and extended emission is much larger than 10. Even during the
late period of the extended emission phase, the $\Delta$BIC of most time slices remains in the
range of [6-10]. By comparing the BIC of 2SBPL and 2SBPL+BB models, we find that it is strong to
support the 2SBPL+BB model, which is better than the 2SBPL model to describe the observed data
during the time-resolved spectra of GRB 211211A. In other words, the thermal emission component in
GRB 211211A remains a significant presence in the time-resolved spectrum. By adopting $z=0.076$ of
GRB 211211A, the total isotropic-equivalent energy ($E_{\rm \gamma,iso}$) and luminosity ($L_{\rm
\gamma,iso}$) can be as high as $7.6\times 10^{51}~\rm erg$ and $1.9\times 10^{51}~\rm erg~s^{-1}$,
respectively.

\subsection{Fitting results}
\begin{figure}
\centering
\includegraphics [angle=0,scale=0.27] {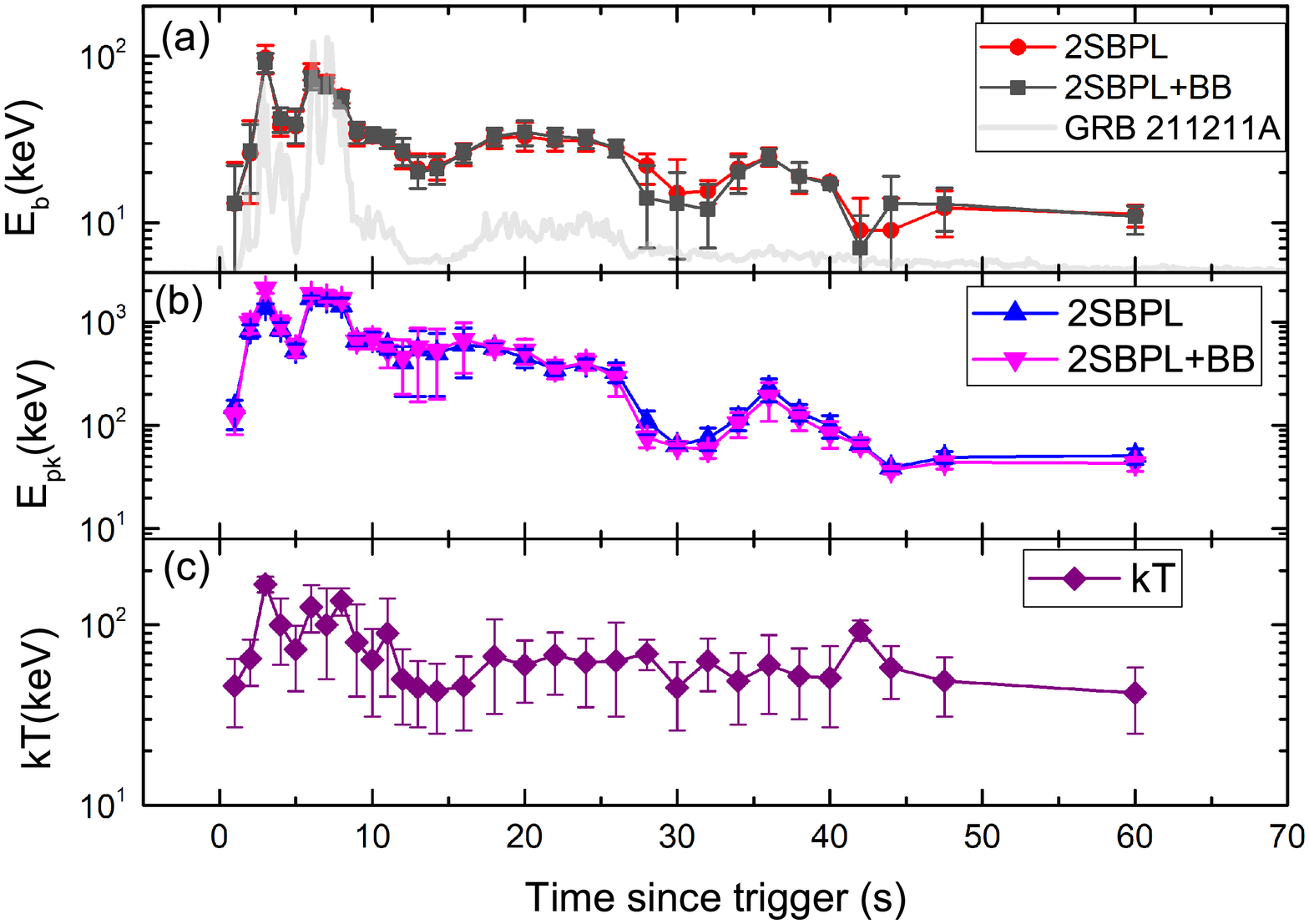}
\includegraphics [angle=0,scale=0.27] {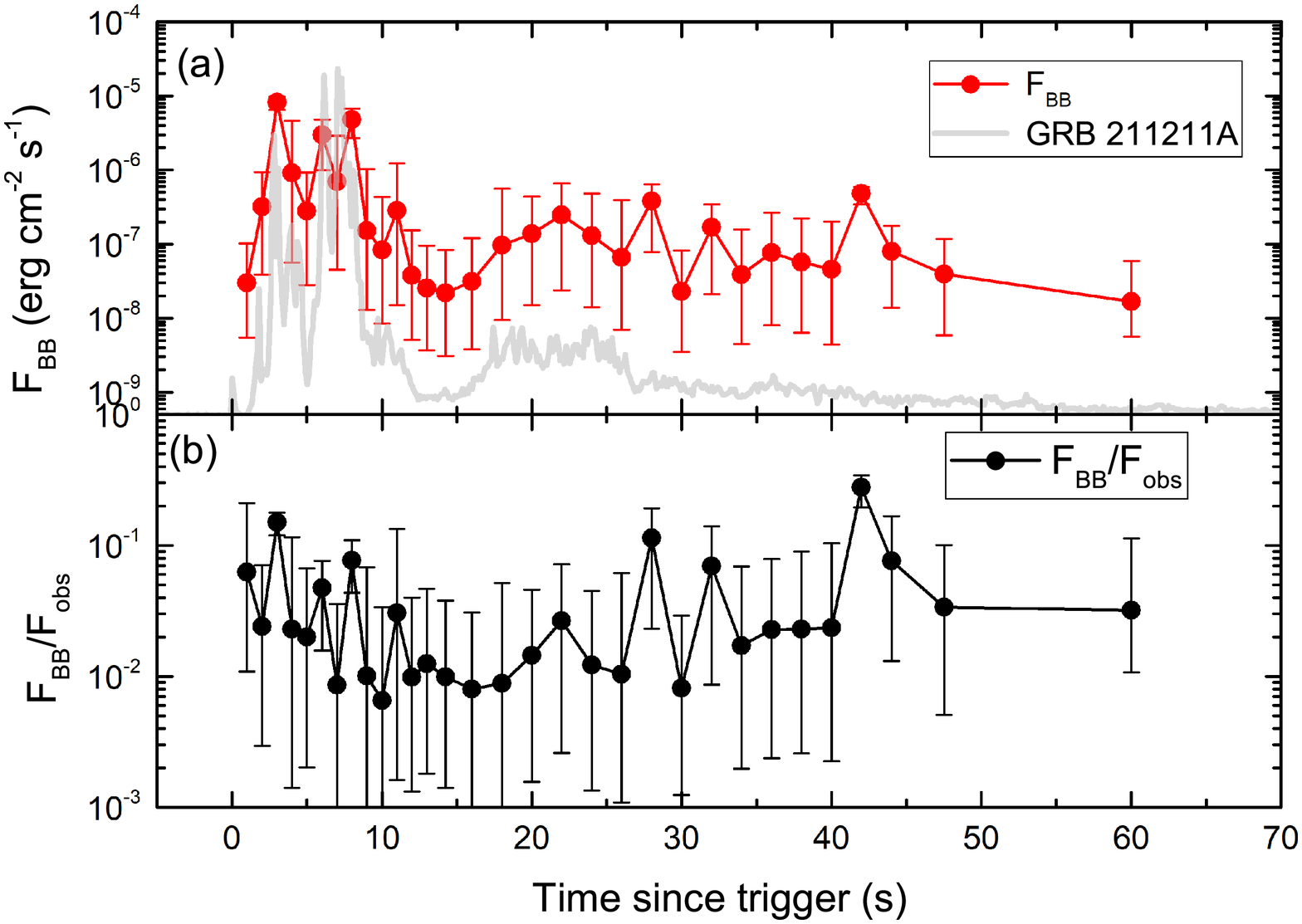}
\caption{Left: temporal evolution of $E_{\rm b}$, $E_{\rm pk}$ and kT of 2SBPL+BB model.
Right: similar to left panels, but for the flux ($F_{\rm BB}$) of BB emission and $F_{\rm BB}/F_{\rm obs}$.}
\label{fig:EpFbb}
\end{figure}

In order to test the behavior of temporal evolution of main parameters for thermal and nonthermal
emissions, we present the temporal evolution of $E_{b}$, $E_{\rm pk}$, $kT$, the flux of BB
emission ($F_{\rm BB}$), and the ratio of $F_{\rm BB}$ and total observed flux ($F_{\rm obs}$) in
Figure \ref{fig:EpFbb}. By comparing the $E_{b}$ and ${E_{\rm pk}}$ evolution of both 2SBPL and
2SBPL+BB models, we find that the $E_{\rm b}$ and $E_{\rm pk}$ evolution of 2SBPL+BB model are
similar to that of the 2SBPL model, and the behavior of its evolution seems to be tracking with
pulses of the light curve. The temperature ($kT$) and the flux ($F_{\rm BB}$) of BB emission also
exhibits the tracking behavior with its pulses. The $F_{\rm BB}$ can reach to as high as
$8.16\times 10^{-6} \rm ~erg~cm^{-2}~s^{-1}$. We also calculate the total observed flux, which
includes both thermal and nonthermal emission and find that the fraction of thermal emission flux
($F_{\rm BB}/F_{\rm obs}$) can reach as high as $\sim 0.2$. The significant thermal component in
the prompt emission is also independent to support the BB emission, which should indeed be
presented.


\section{Derivation of the physical parameters within the Fireball Models}
The initial Lorentz factor of a GRB jet is a very important parameter for understanding GRB
physics, and is also very difficult to measure for most GRBs \citep{2018pgrb.book.....Z}. In
general, there are three methods that have been proposed to estimate the Lorentz factor. The first
one is to use the high-energy cutoff of the prompt gamma-ray spectrum, which is from the pair
production when the absorption optical depth is close to one
\citep{1993A&AS...97...59F,2001ApJ...555..540L,2009ApJ...700L..65Z}. The second approach to
estimating the Lorentz factor is using the early afterglow light curves with a smooth onset bump
that shows the signal of fireball deceleration
\citep{1999A&AS..138..537S,2007ApJ...655..973K,2010ApJ...725.2209L}. The third one is using the
blackbody component detected in some GRB spectra
\citep{2007ApJ...664L...1P,2010ApJ...709L.172R,2015ApJ...801...2}. In this section, we derive the
physical parameters based on the observed thermal component in prompt emission of GRB 211211A, such
as the Lorentz factor ($\Gamma_{\rm ph}$) and the radius of the photosphere ($R_{\rm ph}$). We also
calculate the magnetization factor ($\sigma_0$) and dimensionless entropy ($\eta$) by assuming the
hybrid jet of GRB 211211A.

\subsection{Lorentz Factor and Photosphere Radius}
In our analyses, the time-resolved spectra of the GRB 211211A prompt emission is composed of a
thermal component (BB component) and a nonthermal component 2SBPL component). Following the method
of \citet{2007ApJ...664L...1P}, we estimate the $\Gamma_{\rm ph}$ and $R_{\rm ph}$ with the BB
component derived from our spectral fits in different time slices;
\begin{equation}
\Gamma_{\mathrm{ph}}=\left[(1.06)(1+z)^{2} D_{\mathrm{L}} \frac{Y \sigma_{\mathrm{T}} F_{\mathrm{obs}}}{2
m_{\mathrm{p}} c^{3} \Re}\right]^{1 / 4}
\end{equation}
where $D_{\rm L}$ is the luminosity distance, $m_{\rm p}$ is the proton mass, $\sigma_{\mathrm{T}}$
is the Thomson scattering cross section, and $F_{\rm obs}$ is the observed total flux. In our
calculation, we fix the $Y=1$, which is the ratio between the total fireball energy and the energy
emitted in the $\gamma$-ray. The definition of $\Re$ is written as
\begin{equation}
\Re \equiv (\frac{F_{BB}}{\sigma T^{4}})^{1/2}
\end{equation}
where $\sigma$ and $F_{\rm BB}$ are the Stefan's constant and the observed blackbody component
flux, respectively. On the other hand, the $R_{\rm ph}$ can be expressed as
\begin{equation}
R_{p h}=\left(L \sigma_{T} / 8 \pi \Gamma^{3}_{ph} m_{p} c^{3}\right)
\end{equation}
where $L=4 \pi D_{L}^{2}F_{\rm obs}$ is the luminosity that is measured for bursts with known
redshift. The calculation results are shown in Table 3.

\begin{figure}
\centering
\includegraphics [angle=0,scale=0.4] {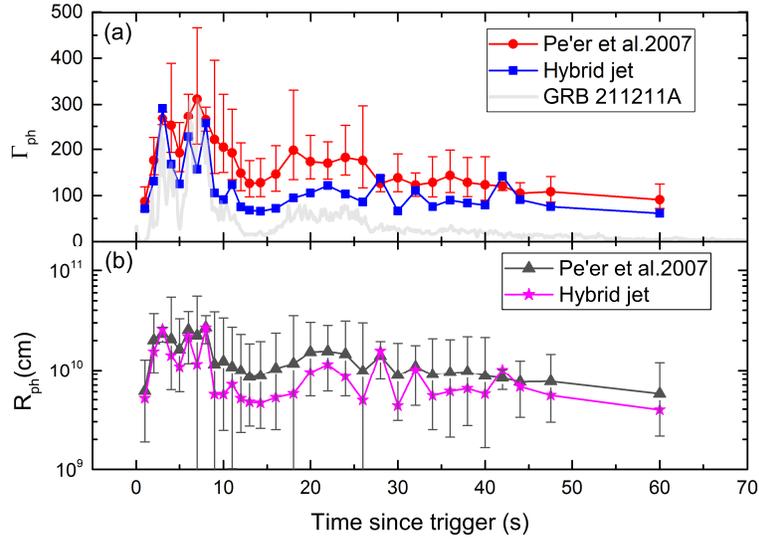}
\caption{Temporal evolution of derived parameters $\Gamma_{\rm ph}$ and $R_{\rm ph}$ by
using the Pe'er et al. (2007) model and hybrid jet (Gao \& Zhang 2015).}
\label{fig:GammaR}
\end{figure}

Figure \ref{fig:GammaR} shows the temporal evolution of $\Gamma_{\rm ph}$ and $R_{\rm ph}$. During
main emission phase, the evolution of $\Gamma_{\rm ph}$ is initially tracking with the light curve,
and maximum value can be reached as high as 311 at the peak of the light curve. In the extended
emission phase, the $\Gamma_{\rm ph}$ seems to be also tracking with the light curve, and it peaked
at $\Gamma_{\rm ph}=197$, then, it is gradually going down to 91 until it reaches the end of
extended emission. As for the $R_{\rm ph}$, the highest value of $R_{\rm ph}$ is around
$\sim2.71\times10^{10}$ cm and it keeps fluctuating around $\sim 10^{10}$ cm.

\subsection{Magnetization parameter and dimensionless entropy}
One is different from the method of \citet{2007ApJ...664L...1P} who inferred the central engine
parameters by using the observed data within the framework of a pure fireball shock model.
\citet{2015ApJ...801...2} proposed a hybrid relativistic outflow of GRB (e.g., hot fireball
component and Poynting-flux component), and developed a theory of its photosphere emission. One
interesting question is that the observed nonthermal component of GRB 211211A is from the internal
shock of a fireball or another cold Poynting-flux component. Here, we infer the magnetization
factor ($\sigma_0$) and dimensionless entropy ($\eta$) by assuming the hybrid jet of GRB 211211A.

The magnetization factor $\sigma_{0}$ is defined as $\sigma_{0}=L_{\rm c}/L_{\rm h}$, where the
$L_{\rm h}$ and $L_{\rm c}$ are the luminosity of the hot fireball component and cold Poynting-flux
component, respectively. The dimensionless entropy $\eta$ can be defined as $\eta=L_{\rm
h}/\dot{M}c^2$, where $\dot{M}$ is the accretion rate. The time varying of the ($\eta$,
$\sigma_{0}$) pair at the central engine can result in the evolution of the photosphere emission
proprieties. Based on the results of derivation in Gao \& Zhang (2015), several situations of
different ($\eta$, $\sigma_{0}$) pairs are considered; (1) ${\eta \gg 1, \sigma_{0} \ll 1}$: it
means that the photosphere emission is dominated by a pure fireball component; (2) ${\eta \sim 1,
1+\sigma_{0} \gg 1}$: a Poynting-flux-dominated outflow; no detection of any thermal component in
the GRB spectrum. By invoking the 'top-down' approach and based on the judgment criteria proposed
by \citet{2015ApJ...801...2}, we can infer the parameters of the central engine by using the
observed quasi-thermal photosphere emission parameters (such as $F_{\rm BB}$, $F_{\rm obs}$, and
$kT$). However, the inferred parameters of the central engine are sensitively dependent on the
selected initial radius ($r_0$) of the central engine. For convenience, we adopt the initial radius
$r_0=10^{7}\rm~cm$, and one can obtain all the photosphere characteristic parameters of the hybrid
model (e.g., $\eta$, $1+\sigma_{0}$, $R_{\rm ph}$, $\Gamma_{\rm ph}$, and $1+\sigma_{\rm ph}$).
Here, $1+\sigma_{\rm ph}$ is the magnetization parameter at $R_{\rm ph}$. Table 4 shows the derived
parameters of GRB 211211A in a hybrid jet model.

\begin{figure}
\centering
\includegraphics [angle=0,scale=0.4] {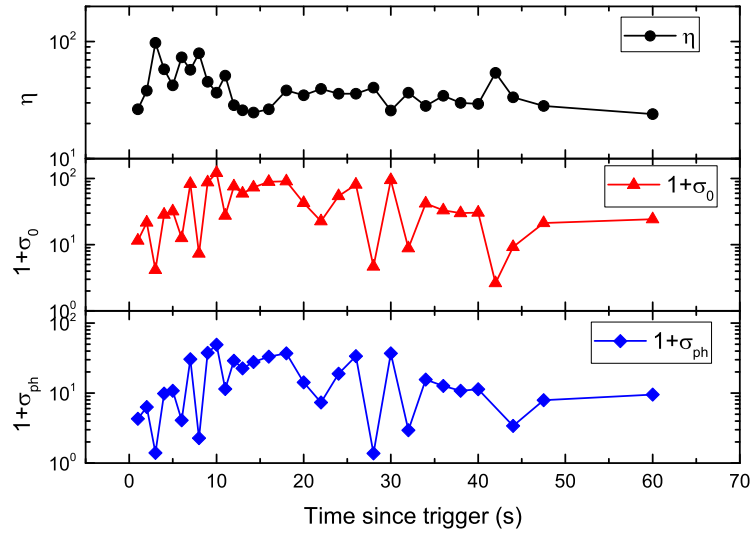}
\caption{Temporal evolution of $1+\sigma_{0}$, $\eta$, and $1+\sigma_{\rm ph}$
in the hybrid jet model for fixed $r_{0}=10^7$ cm.}
\label{fig:sigma}
\end{figure}

Figure \ref{fig:GammaR} also shows the comparisons of $\Gamma_{\rm ph}$ and $R_{\rm ph}$ evolution
for both the pure fireball model and hybrid jet model with fixed $r_0=10^{7}\rm~cm$. It is
interesting that the $\Gamma_{\rm ph}$ evolution behavior of a pure fireball model is similar to
that of the hybrid jet model, and the evolution of $R_{\rm ph}$ is matched very well between those
two models. Figure \ref{fig:sigma} presents the temporal evolution of $1+\sigma_{0}$ and $\eta$ in
the central engine with $r=r_0$, as well as $1+\sigma_{\rm ph}$ at $R_{\rm ph}$. The values of
$1+\sigma_{\rm 0}$ and $\eta$ are larger than 1 and 10 for different time slices, respectively. It
means that at least the Poynting-flux component is indeed existent in the central engine.
Meanwhile, it is clear to see that the values of $1+\sigma_{\rm ph}$ are larger than 1 for
different time slices, and range in [1-50]. So that, those results suggest that the Poynting-flux
component should always be a presence at the position of the central engine and photosphere radius.
The observed thermal and nonthermal components in GRB 211211A seem to be from the contributions of
hot fireball and Poynting-flux outflow, respectively.

\subsection{$\Gamma_{\rm ph}$ - $E_{\gamma,\rm iso}$/$L_{\gamma,\rm iso}$ correlations}
\citet{2010ApJ...725.2209L} discovered a tight correlation between initial Lorentz factor
$\Gamma_0$ and isotropic $\gamma$-ray energy $E_{\rm \gamma,iso}$, namely $\Gamma_0\propto
E^{0.25}_{\rm \gamma,iso}$. \citet{2012ApJ...751...49L} also found another tight correlation of
$\Gamma_0$ and isotropic $\gamma$-ray luminosity $L_{\rm \gamma,iso}$, e.g., $\Gamma_0 \propto
L_{\rm \gamma, iso}^{0.3}$. Those two correlations can be interpreted well by using a
neutrino-cooled hyperaccretion disk around a stellar mass black hole as the GRB central engine
\citep{2012ApJ...751...49L}.

\begin{figure}
\centering
\includegraphics [angle=0,scale=0.27] {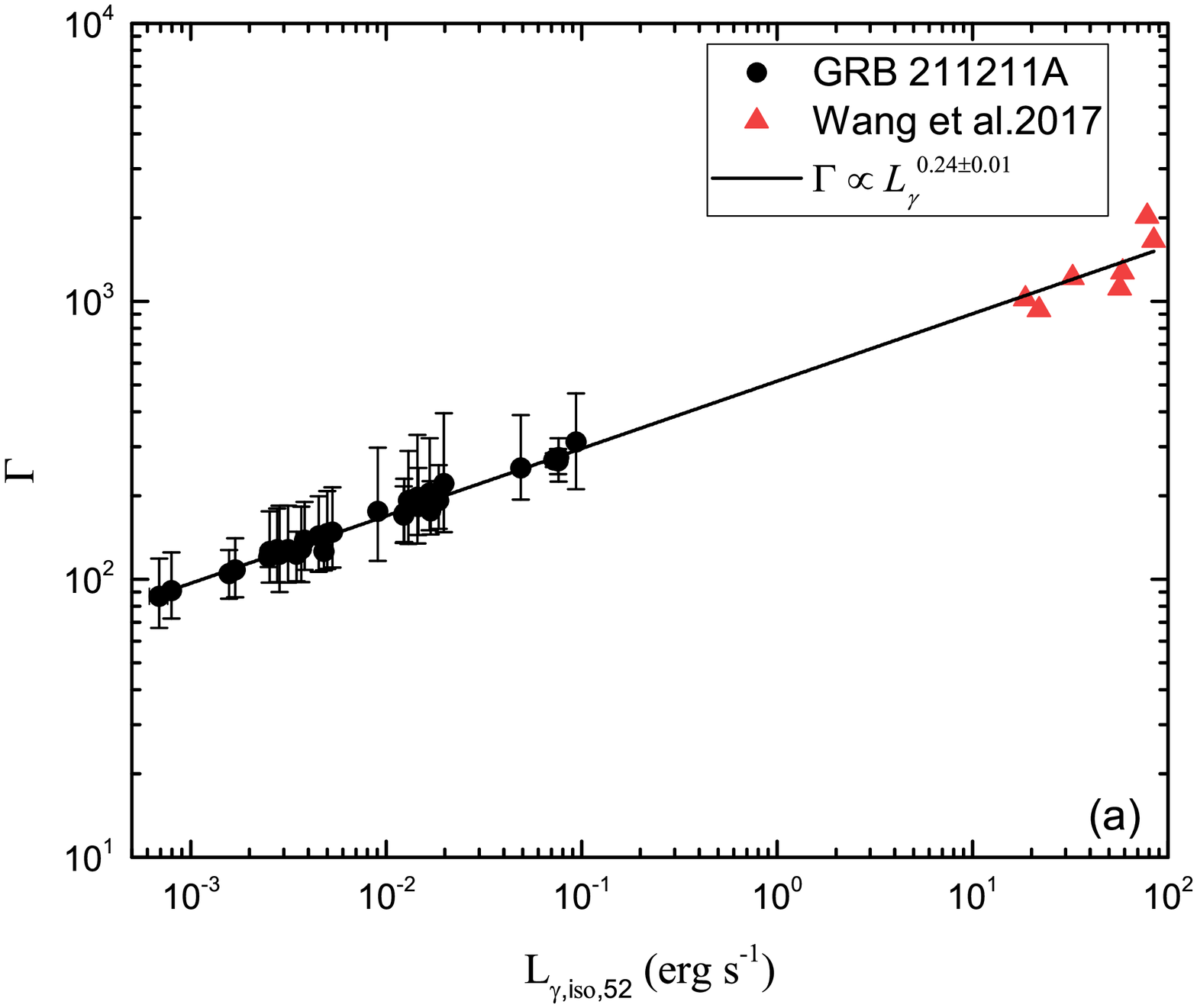}
\includegraphics [angle=0,scale=0.27] {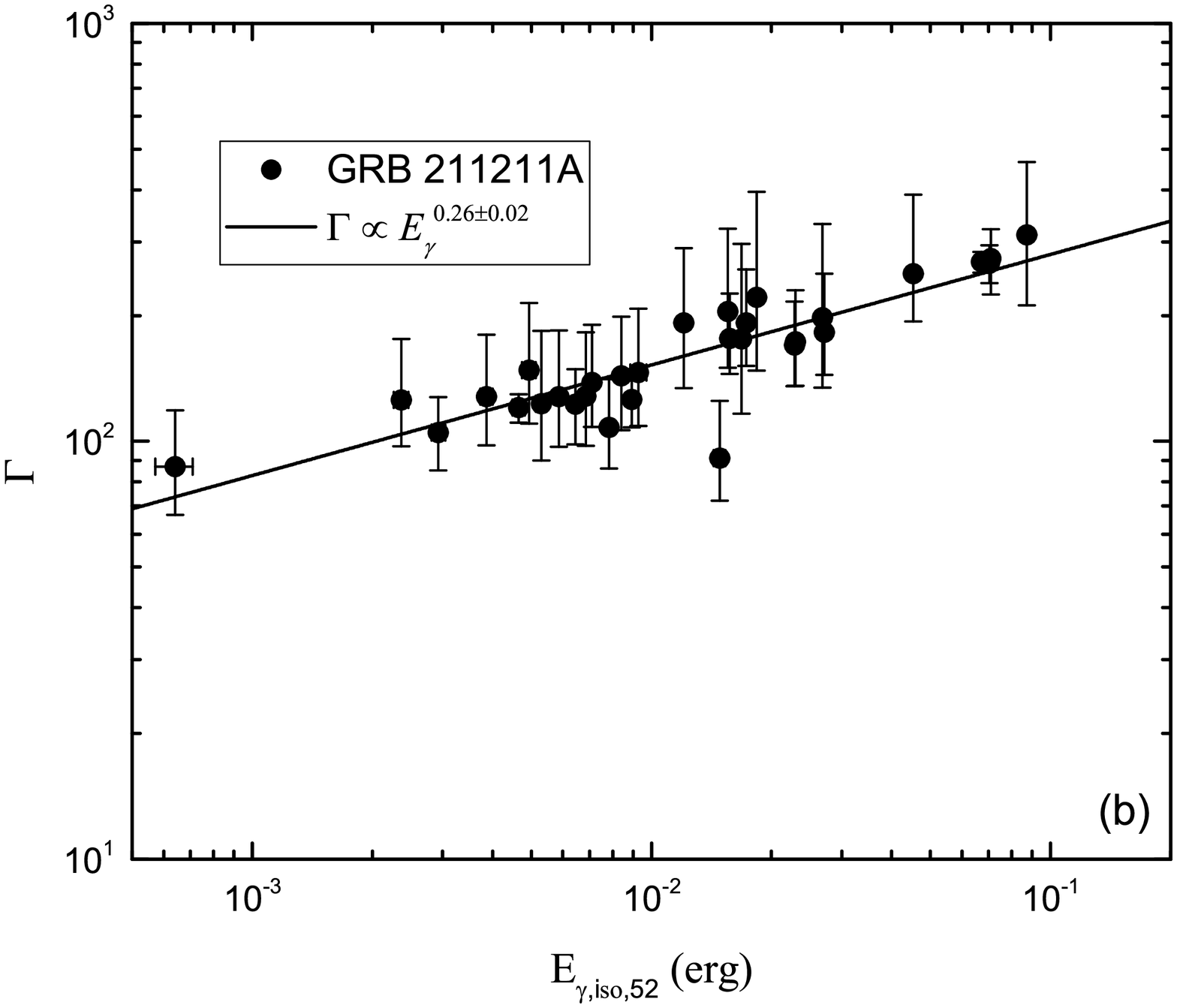}
\caption{Lorentz Factor ($\Gamma_{\rm ph}$) as a function of $L_{\gamma, \rm iso}$ (a) and $E_{\gamma,\rm
iso}$
of GRB 211211A (black solid circles). The black solid lines are the best fit with power-law model. The
red triangles are the data of GRB 160625B from Wang et al. (2017).}
\label{fig:correlations}
\end{figure}

In order to test whether the $\Gamma_{\rm ph}$ from time-resolved BB emission and $E_{\rm
\gamma,iso}$/$L_{\rm \gamma,iso}$ of GRB 211211A are tracking similar correlations to the above,
Figure \ref{fig:correlations} shows the relationship between $\Gamma_{\rm ph}$ and $E_{\rm
\gamma,iso}$/$L_{\rm \gamma,iso}$. We caution the reader that we do not adopt the data from
\citet{2010ApJ...725.2209L} and \citet{2012ApJ...751...49L} to do the joint fitting, because they
adopt different methods to estimate the Lorentz factor. However, we collect the data of
time-resolved spectra in GRB 160625B, which have the thermal emission and use the same method
(e.g., BB emission) to infer the Lorentz factor. So, we used $\Gamma_{\rm ph}$ in our fitting to
replace $\Gamma_{0}$ in \citet{2010ApJ...725.2209L}, and the data of GRB 160625B are taken from
\citet{2017ApJ...836...81W}. For the diagram of $\Gamma_{\rm ph}-L_{\gamma,\rm iso}$, we make a
joint fitting of the two groups (GRB 160626B and GRB 211211A) with a power-law model and find
$\Gamma_{\rm ph}\propto L_{\rm \gamma, iso}^{0.24 \pm 0.01}$ with a Pearsons correlation
coefficient of 0.96 and p $<10^{-4}$. Also, applying the power-law fitting to $\Gamma_{\rm
ph}-E_{\rm \gamma, iso}$, we find $\Gamma_{\rm ph}\propto E_{\rm \gamma, iso}^{0.26 \pm 0.02}$ with
a Pearsons correlation coefficient of 0.91 and p $<10^{-4}$. Those two results are similar to that
of \citet{2012ApJ...751...49L} and \citet{2010ApJ...725.2209L}, respectively. It suggests that the
central engine of GRB 211211A may be a stellar mass black hole with a neutrino-cooled
hyperaccretion disk.


\section{Conclusion and discussion}
GRB 211211A was observed by Fermi/GBM, Swift/BAT, and Insight-HXMT to have a duration of $\sim 84$
seconds at redshift $z=$0.076, but the light curve is characterized by an initial hard-main
emission (with a duration $\sim 13$ s) followed by a series of soft gamma-ray extended emission
with a duration $\sim 55$ s. The structure of the light curve is similar to the cases of GRB 060614
and GRB 211227A, which are believed to be from the compact star merger
\citep{yang2022,xiao2022,2022ApJ...931L..23L}. The total isotropic-equivalent energy ($E_{\rm
\gamma,iso}$) and luminosity ($L_{\rm \gamma,iso}$) are as high as $7.6\times 10^{51}~\rm { erg}$
and $1.9\times 10^{51}~\rm {erg~s^{-1}}$, respectively. At such low redshift, it is surprising that
deep searches of an underlying SN give null results, but being associated with a kilonova is
observed by several optical telescopes \citep{Rastinejad2022}. That observed evidence suggests that
GRB 211211A is originated from a binary compact star merger
\citep{Rastinejad2022,yang2022,xiao2022,Gompertz2022}.

By reanalyzing the data observed with the GBM on board the Fermi mission, we find the following
interesting results:
\begin{itemize}
\item We find that the 2SBPL+BB model is the best one to adequately describe the observed data
    in both time-integrated and time-resolved spectra in the prompt emission of GRB 211211A
    based on the BIC criterion, and it means that both thermal and nonthermal components in GRB
    211211A should be a significant presence.
\item The behavior of temporal evolution of $E_{\rm b}$, $E_{\rm pk}$ for the 2SBPL function
    and $kT$ for BB emission seem to be tracking with pulses of the light curve.
\item By inferring the Lorentz factor $\Gamma_{\rm ph}$ and photosphere radius $R_{\rm ph}$
    based on the observed BB emission within the framework of a pure fireball model, we find
    that the temporal evolution of $\Gamma_{\rm ph}$ seems to be tracking with the light curve,
    and its range from 87 to 311. However, the highest value of $R_{\rm ph}$ is around
    $\sim2.7\times10^{10}$ cm and it keeps fluctuating around $\sim 10^{10}$ cm.
\item By calculating the magnetization factor $\sigma_{0}$ in the central engine and
    $\sigma_{\rm ph}$ at the photosphere radius within the framework of the hybrid jet model,
    we find that the values of both $1+\sigma_{\rm 0}$ and $1+\sigma_{\rm ph}$ are larger than
    1 for different time slices for fixed initial $r_0=10^{7}\rm~cm$, and its range of [1,120].
    It means that at least the Poynting-flux component is indeed existent in both the central
    engine and photosphere radius.
\item Moreover, we also plot the diagrams of $\Gamma_{\rm ph}$- $E_{\rm \gamma,iso}$/$L_{\rm
    \gamma,iso}$, and find the relationships as $\Gamma_{\rm ph}\propto E_{\rm \gamma,
    iso}^{0.26 \pm 0.02}$ and $\Gamma_{\rm ph}\propto L_{\rm \gamma, iso}^{0.24 \pm 0.01}$ for
    the time-resolved spectral data of GRB 211211A. Those two correlations are consistent with
    that of in \citet{2010ApJ...725.2209L} and \citet{2012ApJ...751...49L}, respectively. It
    suggests that the central of GRB 211211A may be a stellar mass black hole with a
    neutrino-cooled hyperaccretion disk.
\end{itemize}

If the central engine of GRB 211211A is a stellar mass black hole that is formed from a binary
compact star merger \citep{Gao2022}. \citet{2017ApJ...849...47L} proposed that two jet launching
mechanisms, i.e., $\nu\bar{\nu}$ annihilation and the Blandford-Znajek (BZ;
\citealt{1977MNRAS.179..433B}) process, are indeed considered to power thermal and nonthermal
components in the relativistic jet, respectively; the $\nu\bar{\nu}$ annihilation mechanism
liberates the gravitational energy from the accretion disk, and the BZ mechanism extracts the spin
energy from the Kerr black hole
\citep{1999ApJ...518..356P,2000PhR...325...83L,2000ApJ...534L.197L,2006ApJ...643L..87G,2009ApJ...700.1970L,2013ApJ...765..125L}.
In this scenario, the observed thermal and nonthermal emissions in GRB 211211A can be interpreted
as follows. The jet of GRB 211211A can be launched from a hyperaccreting black hole via
$\nu\bar{\nu}$ annihilation and the BZ mechanism. Initially, due to a very high accretion rate of
the black hole, the $\nu\bar{\nu}$ annihilation should be dominated. It can produce thermal
emission when the photons escape the system at the photosphere radius, and the internal shock or
Poynting-flux outflow is used to produce the nonthermal component. After tens of seconds, the
$\nu\bar{\nu}$ annihilation is not strong enough along with the decreasing of accretion rate. But
the magnetic field of the black hole becomes gradually stronger due to storage time, and the BZ
mechanism will be dominated. If this is the case, it is natural to explain the observed gradually
decreased thermal emission and gradually increased nonthermal emission of GRB 211211A.

Alternatively, what we discussed above about the evidence of Poynting-flux existence (e.g.,
magnetization factor $\sigma_{0}$) in our results is dependent on the selected $R_0$ in the
calculations of the hybrid model, and we fixed $R_0=10^7$ cm in our calculations. For different
selected $R_0$, such as $R_0=10^8$ cm or $R_0=10^9$ cm, it corresponds to different values of
$\sigma_{0}$. So, it is difficult to judge whether it is accompanied by the Poynting-flux component
in the jet. If this is the case, only $\nu\bar{\nu}$ annihilation mechanism can interpret the
observed thermal emission (from the photosphere) and nonthermal emission (from the internal shock).

To find out such a uniform model to explain all observed characteristics of GRB 211211A, of course,
is not an easy task. We therefore encourage intense multi-band follow-up observations for GRB
211211A-like events in the future.

\begin{acknowledgements}
We are very grateful to thank the referee for helpful comments and suggestions to improve this
manuscript. We acknowledge the use of the public data from the Fermi/GBM data archive. This work is
supported by the National Natural Science Foundation of China (grant Nos. 11922301, and 12133003),
the Program of Bagui Young Scholars Program (LHJ), and the Guangxi Science Foundation (grant No.
2017GXNSFFA198008).

\end{acknowledgements}



\begin{table*} \centering
\renewcommand\tabcolsep{2.2pt}
\renewcommand\arraystretch{1.4}
\caption{Time-integral spectral analysis results of GRB 211211A.}
\begin{tabular}{cccccccccc} 
\hline\hline
$t_{1}-t_{2}$&$\rm Model$&$\alpha(\alpha_{1})$&$E_{\rm b}$&$E_{\rm p}/E_{\rm
c}$&$\beta$&$\alpha_{2}$&kT&BIC&\rm favorite Model \\
(s)&$\quad$&$\quad$&(keV)&(keV)&$\quad$&$\quad$&(keV)&$\quad$&$\quad$\\
\hline
0.5-70	&	Band	&$-1.22^{+0.01}_{-0.01}$&...&	$497^{+15}_{-15}$&	$-2.21^{+0.03}_{-0.03}$&...	&...&	
8633	&	$\quad$\\
$\quad$&	CPL	&$-1.27^{+0.01}_{-0.01}$&...&	$916^{+19}_{-19}$&	...&...&...&	8964	&	$\quad$\\
$\quad$&	BB	&...&	...&	...&...&...&	$~53$&	125651	&	$\quad$\\
$\quad$&	Band+BB	&$-1^{+0.01}_{-0.01}$&...&	$175^{+6}_{-6}$&	$-2.06^{+0.02}_{-0.02}$&...&	
$201^{+7}_{-7}$&	7864	&	$\quad$\\
$\quad$&	CPL+BB	&$-1.08^{+0.01}_{-0.01}$&...&	$250^{+5}_{-5}$&...&...&	$312^{+5}_{-5}$	&8149	&	
$\quad$\\
$\quad$&	SBPL	&$-1.19^{+0.01}_{-0.01}$&$210^{+10}_{-11}$&...&${-2.11^{+0.02}_{-0.02}}$ &...&	...&	
8523	&	$\quad$\\
$\quad$&    SBPL+BB  &$-0.96^{+0.02}_{-0.02}$&$81^{+4}_{-4}$&...&${-2.08^{+0.02}_{-0.02}}$ &...&
$199^{+6}_{-6}$ &	7732	& $\quad$\\
$\quad$&
2SBPL&${-0.6^{+0.04}_{-0.04}}$&$26.2^{+1.3}_{-1.3}$&$1020^{+40}_{-50}$&$-2.93^{+0.07}_{-0.07}$&$-1.59^{+0.01}_{-0.01}$&...&7461&$\quad$\\
$\quad$& 2SBPL+BB
&${-0.6^{+0.04}_{-0.04}}$&$27^{+1.9}_{-1.8}$&$1200^{+110}_{-110}$&$-3.14^{+0.14}_{-0.13}$&$-1.62^{+0.02}_{-0.02}$&$77^{+16}_{-15}$&7443&$\surd$\\
\hline

\end{tabular}
\label{fit_data}
\end{table*}


\begin{sidewaystable}
\centering
\renewcommand\tabcolsep{0.2pt}
\renewcommand\arraystretch{0.1}
\caption{The fitting results of time-resolved spectra with 2SBPL and 2SBPL+BB in GRB 211211A prompt
emission.}
\begin{tabular}{cccccccccccccccc} 
\hline\hline
$t_{1}$ & $t_{2}$ &$\quad$ &$\quad$& 2SBPL &$\quad$&$\quad$ &BIC (PGSTAT)&$\quad$&$\quad$&2SBPL +
BB&$\quad$&$\quad$&$\quad$&BIC (PGSTAT)&$\Delta$BIC\\
s & s & $\alpha_{1}$ &$\alpha_{2}$&$E_{\rm b}$(keV) &$E_{\rm pk}$(keV) & $\beta$ & $\quad$ & $\alpha_{1}$
&$\alpha_{2}$&$E_{\rm b}$(keV) & $E_{\rm pk}$(keV) & $\beta$
& kT(keV) & $\quad$ &$\quad$\\
\hline
0.5	&	1.5	&$	-0.9^{+	0.5	}_{-	0.5	}$&$	-1.1^{+	0.6	}_{-0.5	}$&$	13^{+	10	}_{-	9	}$&$	
151^{+	24	}_{-	60	}$&$	-2.9^{+	0.5	}_{-	0.5	}$&

1640.4(1605.2)&$	-0.9^{+	0.5	}_{-	0.4	}$&$	-1.1^{+	0.5	}_{-	0.5	}$&$	13^{+	9	}_{-	9	 
}$&$	121^{+	24	}_{-	40	}$&$	-3.1^{+	0.5	}_{-

0.5	}$&$	46^{+	19	}_{-	19	}$&	1632.6(1585.7)&	7.8	\\
1.5	&	2.5	&$	-0.9^{+	0.4	}_{-	0.5	}$&$	-1.07^{+	0.4	}_{-0.33	}$&$	26^{+	15	}_{-	13	 
}$&$	820^{+	120	}_{-	120	}$&$	-2.52^{+	0.14

}_{-	0.15	}$&	2001.2(1966.1)&$	-0.83^{+	0.34	}_{-	0.5	}$&$	-1.17^{+	0.5	}_{-	0.3	 
}$&$	27^{+	12	}_{-	12	}$&$	980^{+	210	}_{-

210	}$&$	-2.66^{+	0.22	}_{-	0.22	}$&$	65^{+	18	}_{-	19	}$&	1999.9(1953.1)&	1.3	\\
2.5	&	3.5	&$	-0.73^{+	0.9	}_{-	0.01	}$&$	-1.48^{+	0.1	}_{-0.1	}$&$	98^{+	18	}_{-	 
19	}$&$	1370^{+	130	}_{-	130	}$&$	-3.2^{+	0.16

}_{-	0.17	}$&	2594.7(2559.5)&$	-0.67^{+	0.04	}_{-	0.04	}$&$	-1.64^{+	0.04	}_{-	 
0.04	}$&$	92^{+	12	}_{-	12	}$&$	2100^{+

210	}_{-	210	}$&$	-3.99^{+	0.31	}_{-	0.31	}$&$	168^{+	17	}_{-	16	}$&	 
2569.5(2522.6)&	25.2	\\
3.5	&	4.5	&$	-0.6^{+	0.4	}_{-	0.8	}$&$	-1.09^{+	0.8	}_{-0.34	}$&$	38^{+	5	}_{-	5	 
}$&$	840^{+	150	}_{-	100	}$&$	-2.88^{+	0.09

}_{-	0.09	}$&	2327.8(2292.6)&$	-0.61^{+	0.33	}_{-	0.8	}$&$	-1.21^{+	0.8	}_{-	0.35	 
}$&$	42^{+	7	}_{-	7	}$&$	960^{+	190	
}_{-	190	}$&$	-2.99^{+	0.18	}_{-	0.14	}$&$	100^{+	40	}_{-	40	}$&	2326(2279.1)&	 
1.8	\\
4.5	&	5.5	&$	-0.68^{+	0.13	}_{-	0.12	}$&$	-1.66^{+	0.06	}_{-0.08	}$&$	38^{+	9	 
}_{-	9	}$&$	530^{+	80	}_{-	80	}$&$	
-3.2^{+	0.4	}_{-	0.4	}$&	2102.9(2067.7)&$	-0.69^{+	0.12	}_{-	0.11	}$&$	-1.71^{+	0.08	 
}_{-	0.08	}$&$	39^{+	9	}_{-	9	}$&$

570^{+	110	}_{-	110	}$&$	-3.3^{+	0.4	}_{-	0.4	}$&$	73^{+	26	}_{-	30	}$&	2095.1(2048.2)&	 
7.8	\\
5.5	&	6.5	&$	-0.62^{+	0.04	}_{-	0.03	}$&$	-1.55^{+	0.02	}_{-0.03	}$&$	81^{+	9	 
}_{-	9	}$&$	1700^{+	100	}_{-	100	}$&$	
-3.36^{+	0.14	}_{-	0.14	}$&	2592.3(2557.2)&$	-0.59^{+	0.05	}_{-	0.04	}$&$	 
-1.55^{+	0.24	}_{-	0.04	}$&$	72^{+	9	}_{-

9	}$&$	1860^{+	170	}_{-	160	}$&$	-3.48^{+	0.17	}_{-	0.18	}$&$	126^{+	40	}_{-	 
35	}$&	2579.9(2533.1)&	12.4	\\
6.5	&	7.5	&$	-0.57^{+	0.12	}_{-	0.03	}$&$	-1.35^{+	0.05	}_{-0.1	}$&$	70^{+	7	 
}_{-	7	}$&$	1630^{+	70	}_{-	110	}$&$	
-3.09^{+	0.09	}_{-	0.09	}$&	2751.6(2716.4)&$	-0.69^{+	0.25	}_{-	0.7	}$&$	-1.22^{+	 
0.7	}_{-	0.24	}$&$	67^{+	7	}_{-	7	

}$&$	1750^{+	280	}_{-	190	}$&$	-3.12^{+	0.1	}_{-	0.1	}$&$	100^{+	60	}_{-	50	}$&	 
2743.6(2696.7)&	8	\\
7.5	&	8.5	&$	0.68^{+	0.29	}_{-	0.8	}$&$	-1.21^{+	0.8	}_{-0.28	}$&$	57^{+	5	}_{-	5	 
}$&$	1440^{+	270	}_{-	160	}$&$	-3.34^{+	
0.13	}_{-	0.31	}$&	2539.6(2504.4)&$	-0.56^{+	0.2	}_{-	0.05	}$&$	-1.35^{+	0.09	 
}_{-	0.2	}$&$	55^{+	6	}_{-	6	}$&$	1670^{+

180	}_{-	180	}$&$	-3.64^{+	0.21	}_{-	0.22	}$&$	136^{+	24	}_{-	23	}$&	 
2524.6(2477.7)&	15	\\
8.5	&	9.5	&$	-0.5^{+	0.11	}_{-	0.11	}$&$	-1.67^{+	0.04	}_{-0.05	}$&$	34^{+	5	 
}_{-	5	}$&$	660^{+	90	}_{-	90	}$&$	-3.2^{+

0.35	}_{-	0.4	}$&	2090.2(2055.1)&$	-0.51^{+	0.11	}_{-	0.11	}$&$	-1.69^{+	0.05	 
}_{-	0.06	}$&$	35^{+	5	}_{-	5	}$&$	
650^{+	120	}_{-	100	}$&$	-3.2^{+	0.4	}_{-	0.4	}$&$	80^{+	50	}_{-	40	}$&	2080.5(2033.7)&	 
9.7	\\
9.5	&	10.5	&$	-0.33^{+	0.11	}_{-	0.11	}$&$	-1.75^{+	0.04	}_{-0.04	}$&$	 
33.4^{+	3.2	}_{-	3.2	}$&$	680^{+	120	}_{-	130	
}$&$

-3.2^{+	0.4	}_{-	0.4	}$&	2072.5(2037.3)&$	-0.34^{+	0.11	}_{-	0.11	}$&$	-1.76^{+	0.04	 
}_{-	0.04	}$&$	33.8^{+	3.4	}_{-	3.4	}$&$

690^{+	160	}_{-	140	}$&$	-3.2^{+	0.4	}_{-	0.4	}$&$	64^{+	31	}_{-	33	}$&	2066.6(2019.8)&	 
5.9	\\
10.5	&	11.5	&$	-0.41^{+	0.12	}_{-	0.12	}$&$	-1.8^{+	0.04	}_{-0.04	}$&$	31^{+	 
3.1	}_{-	3	}$&$	590^{+	90	}_{-	100	
}$&$

-3.9^{+	0.5	}_{-	0.5	}$&	1991.1(1955.9)&$	-0.44^{+	0.12	}_{-	0.12	}$&$	-1.85^{+	0.07	 
}_{-	0.08	}$&$	32^{+	4	}_{-	4	}$&$

530^{+	150	}_{-	170	}$&$	-3.8^{+	0.5	}_{-	0.5	}$&$	90^{+	50	}_{-	50	}$&	1981.1(1934.3)&	 
10	\\
11.5	&	12.5	&$	-0.68^{+	0.21	}_{-	0.19	}$&$	-1.82^{+	0.07	}_{-0.14	}$&$	 
26^{+	6	}_{-	5	}$&$	410^{+	170	}_{-	220	
}$&$	-3^{+	0.7	}_{-	0.7	}$&	1850.3(1815.1)&$	-0.68^{+	0.16	}_{-	0.17	}$&$	-1.89^{+	 
0.06	}_{-	0.08	}$&$	27^{+	5	}_{-	5

}$&$	450^{+	220	}_{-	250	}$&$	-3.1^{+	0.6	}_{-	0.6	}$&$	50^{+	23	}_{-	22	}$&	 
1843.1(1796.2)&	7.2	\\
12.5	&	13.5	&$	-0.56^{+	0.32	}_{-	0.32	}$&$	-1.83^{+	0.1	}_{-0.12	}$&$	21^{+	 
5	}_{-	5	}$&$	540^{+	280	}_{-	350	}$&$

-2.9^{+	0.6	}_{-	0.6	}$&	1711.8(1676.6)&$	-0.53^{+	0.3	}_{-	0.29	}$&$	-1.86^{+	0.09	 
}_{-	0.11	}$&$	20^{+	5	}_{-	4	}$&$	
570^{+	300	}_{-	400	}$&$	-2.9^{+	0.6	}_{-	0.6	}$&$	45^{+	18	}_{-	18	}$&	1703.9(1657.1)&	 
7.9	\\
13.5	&	15	&$	-0.63^{+	0.23	}_{-	0.25	}$&$	-1.88^{+	0.07	}_{-0.08	}$&$	22^{+	 
4	}_{-	4	}$&$	500^{+	270	}_{-	310	}$&$

-3^{+	0.6	}_{-	0.6	}$&	2154.1(2118.8)&$	-0.65^{+	0.24	}_{-	0.23	}$&$	-1.86^{+	0.05	 
}_{-	0.12	}$&$	21^{+	4	}_{-	4	}$&$

530^{+	320	}_{-	350	}$&$	-3^{+	0.6	}_{-	0.6	}$&$	43^{+	18	}_{-	18	}$&	2152.1(2105.3)&	2	 
\\
15	&	17	&$	-0.53^{+	0.19	}_{-	0.17	}$&$	-1.81^{+	0.05	}_{-0.13	}$&$	26^{+	4	 
}_{-	4	}$&$	600^{+	270	}_{-	310	}$&$	
-2.8^{+	0.5	}_{-	0.6	}$&	2624.1(2588.8)&$	-0.52^{+	0.15	}_{-	0.16	}$&$	-1.88^{+	0.06	 
}_{-	0.07	}$&$	26.1^{+	3.5	}_{-	3.3	}$&$

670^{+	310	}_{-	350	}$&$	-2.9^{+	0.6	}_{-	0.6	}$&$	46^{+	21	}_{-	20	}$&	2618.4(2571.6)&	 
5.7	\\
17	&	19	&$	-0.5^{+	0.1	}_{-	0.1	}$&$	-1.71^{+	0.04	}_{-0.04	}$&$	32^{+	4	}_{-	4	 
}$&$	560^{+	70	}_{-	70	}$&$	-3.31^{+	
0.34	}_{-	0.35	}$&	2860.4(2825.2)&$	-0.52^{+	0.1	}_{-	0.1	}$&$	-1.74^{+	0.05	}_{-	 
0.05	}$&$	33^{+	4	}_{-	4	}$&$	570^{+

90	}_{-	80	}$&$	-3.33^{+	0.4	}_{-	0.35	}$&$	67^{+	40	}_{-	35	}$&	2853.4(2806.6)&	7	 
\\
19	&	21	&$	-0.63^{+	0.13	}_{-	0.13	}$&$	-1.75^{+	0.08	}_{-0.09	}$&$	33^{+	7	 
}_{-	6	}$&$	450^{+	90	}_{-	90	}$&$	
-3^{+	0.4	}_{-	0.4	}$&	2781.7(2746.5)&$	-0.66^{+	0.1	}_{-	0.1	}$&$	-1.81^{+	0.08	}_{-	 
0.08	}$&$	35^{+	6	}_{-	6	}$&$	530^{+

150	}_{-	140	}$&$	-3.2^{+	0.5	}_{-	0.5	}$&$	60^{+	22	}_{-	23	}$&	2774.5(2727.7)&	7.2	\\
21	&	23	&$	-0.57^{+	0.1	}_{-	0.1	}$&$	-1.79^{+	0.05	}_{-0.05	}$&$	31^{+	4	}_{-	 
4	}$&$	350^{+	50	}_{-	50	}$&$	-3.2^{+	
0.31	}_{-	0.32	}$&	2822.7(2787.5)&$	-0.59^{+	0.09	}_{-	0.1	}$&$	-1.84^{+	0.07	 
}_{-	0.08	}$&$	33^{+	4	}_{-	4	}$&$	
350^{+	80	}_{-	70	}$&$	-3.2^{+	0.4	}_{-	0.4	}$&$	68^{+	23	}_{-	27	}$&	2814.3(2767.5)&	 
8.4	\\
23	&	25	&$	-0.52^{+	0.1	}_{-	0.1	}$&$	-1.75^{+	0.05	}_{-0.05	}$&$	31^{+	4	}_{-	 
4	}$&$	390^{+	50	}_{-	50	}$&$	-3.25^{+

0.32	}_{-	0.31	}$&	2836.4(2801.2)&$	-0.53^{+	0.11	}_{-	0.11	}$&$	-1.78^{+	0.06	 
}_{-	0.06	}$&$	32^{+	4	}_{-	4	}$&$

410^{+	80	}_{-	70	}$&$	-3.3^{+	0.4	}_{-	0.4	}$&$	62^{+	22	}_{-	27	}$&	2829.1(2782.2)&	 
7.3	\\
25	&	27	&$	-0.53^{+	0.12	}_{-	0.12	}$&$	-1.86^{+	0.05	}_{-0.05	}$&$	28.1^{+	 
3.3	}_{-	3.4	}$&$	330^{+	70	}_{-	70	}$&$	

-3.6^{+	0.5	}_{-	0.5	}$&	2718.9(2683.7)&$	-0.54^{+	0.12	}_{-	0.12	}$&$	-1.87^{+	0.06	 
}_{-	0.06	}$&$	28.1^{+	3.4	}_{-	3.4	}$&$

290^{+	90	}_{-	100	}$&$	-3.5^{+	0.6	}_{-	0.6	}$&$	63^{+	40	}_{-	32	}$&	2706.8(2660.1)&	 
12.1	\\
27	&	29	&$	-0.56^{+	0.17	}_{-	0.18	}$&$	-1.78^{+	0.2	}_{-0.17	}$&$	22^{+	4	 
}_{-	5	}$&$	108^{+	30	}_{-	27	}$&$	-3.1^{+

0.6	}_{-	0.6	}$&	2643.6(2608.4)&$	-0.7^{+	0.5	}_{-	0.6	}$&$	-1.1^{+	0.7	}_{-	0.6	}$&$	14^{+	 
8	}_{-	7	}$&$	76^{+	11	}_{-	15	}$&$

-2.69^{+	0.25	}_{-	0.19	}$&$	69^{+	14	}_{-	13	}$&	2628.7(2581.9)&	14.9	\\
29	&	31	&$	-0.81^{+	0.32	}_{-	0.3	}$&$	-0.9^{+	0.4	}_{-0.4	}$&$	15^{+	9	}_{-	9	 
}$&$	64^{+	5	}_{-	5	}$&$	-2.29^{+	0.06

}_{-	0.06	}$&	2601.8(2566.6)&$	-0.88^{+	0.4	}_{-	0.31	}$&$	-0.8^{+	0.4	}_{-	0.4	}$&$	 
13^{+	7	}_{-	7	}$&$	62^{+	5	}_{-	5

}$&$	-2.31^{+	0.07	}_{-	0.06	}$&$	45^{+	17	}_{-	19	}$&	2594.1(2547.2)&	7.7	\\
31	&	33	&$	-0.37^{+	0.24	}_{-	0.24	}$&$	-1.81^{+	0.18	}_{-0.15	}$&$	15.5^{+	 
2.4	}_{-	2.5	}$&$	75^{+	19	}_{-	18	}$&$	

-2.8^{+	0.4	}_{-	0.4	}$&	2616.2(2581)&$	-0.8^{+	0.5	}_{-	0.5	}$&$	-1.2^{+	0.7	}_{-	0.7	}$&$	 
12^{+	5	}_{-	5	}$&$	58^{+	9	}_{-	10	
}$&$	-2.6^{+	0.23	}_{-	0.22	}$&$	63^{+	21	}_{-	20	}$&	2605.6(2558.8)&	10.6	\\
33	&	35	&$	-0.45^{+	0.27	}_{-	0.23	}$&$	-1.72^{+	0.14	}_{-0.21	}$&$	21^{+	5	 
}_{-	5	}$&$	116^{+	28	}_{-	27	}$&$	
-3^{+	0.5	}_{-	0.5	}$&	2633.9(2598.7)&$	-0.51^{+	0.32	}_{-	0.24	}$&$	-1.59^{+	0.19	 
}_{-	0.34	}$&$	20^{+	5	}_{-	5	}$&$

105^{+	28	}_{-	29	}$&$	-2.9^{+	0.4	}_{-	0.4	}$&$	49^{+	21	}_{-	21	}$&	2627.4(2580.5)&	 
6.5	\\
35	&	37	&$	-0.56^{+	0.15	}_{-	0.15	}$&$	-1.89^{+	0.05	}_{-0.06	}$&$	24.9^{+	 
3.3	}_{-	3.1	}$&$	230^{+	50	}_{-	60	}$&$	

-3.6^{+	0.6	}_{-	0.6	}$&	2640.5(2605.3)&$	-0.55^{+	0.13	}_{-	0.14	}$&$	-1.91^{+	0.06	 
}_{-	0.06	}$&$	24.9^{+	2.8	}_{-	2.8	}$&$

190^{+	70	}_{-	80	}$&$	-3.4^{+	0.6	}_{-	0.6	}$&$	60^{+	28	}_{-	28	}$&	2631.5(2584.6)&	9	 
\\
37	&	39	&$	-0.53^{+	0.23	}_{-	0.23	}$&$	-1.83^{+	0.08	}_{-0.09	}$&$	19^{+	4	 
}_{-	4	}$&$	135^{+	24	}_{-	24	}$&$	
-3.5^{+	0.5	}_{-	0.5	}$&	2606.6(2571.4)&$	-0.55^{+	0.23	}_{-	0.22	}$&$	-1.82^{+	0.08	 
}_{-	0.11	}$&$	18.9^{+	4	}_{-	3.4	}$&$

119^{+	29	}_{-	30	}$&$	-3.5^{+	0.5	}_{-	0.5	}$&$	52^{+	22	}_{-	22	}$&	2597.9(2551.1)&	 
8.7	\\
39	&	41	&$	-0.29^{+	0.3	}_{-	0.25	}$&$	-1.8^{+	0.07	}_{-0.15	}$&$	17.5^{+	0.26	 
}_{-	0.26	}$&$	99^{+	26	}_{-	24	}$&$	
-3.2^{+	0.5	}_{-	0.6	}$&	2557.7(2522.5)&$	-0.23^{+	0.3	}_{-	0.28	}$&$	-1.81^{+	0.14	 
}_{-	0.13	}$&$	17.1^{+	0.26	}_{-	0.26	
}$&$	84^{+	24	}_{-	24	}$&$	-3.2^{+	0.5	}_{-	0.5	}$&$	51^{+	25	}_{-	24	}$&	 
2547.1(2500.3)&	10.6	\\
41	&	43	&$	-1^{+	0.5	}_{-	0.4	}$&$	-1.1^{+	0.6	}_{-0.4	}$&$	9^{+	5	}_{-	5	}$&$	 
66^{+	8	}_{-	7	}$&$	-2.44^{+	0.08	}_{-

0.15	}$&	2578.5(2543.3)&$	-1.1^{+	0.5	}_{-	0.4	}$&$	-1^{+	0.8	}_{-	0.6	}$&$	7^{+	4	 
}_{-	4	}$&$	65^{+	11	}_{-	9	}$&$	-3.4^{+

0.4	}_{-	0.4	}$&$	93^{+	13	}_{-	11	}$&	2563.5(2516.7)&	15	\\
43	&	45	&$	-0.4^{+	0.4	}_{-	0.4	}$&$	-1.2^{+	0.4	}_{-0.4	}$&$	9^{+	5	}_{-	4	}$&$	 
38.9^{+	3	}_{-	3	}$&$	-2.31^{+	0.18	
}_{-

0.24	}$&	2511.5(2476.3)&$	-0.48^{+	0.35	}_{-	0.4	}$&$	-1.26^{+	0.27	}_{-	0.27	 
}$&$	13^{+	6	}_{-	8	}$&$	37^{+	3.2	}_{-

3.1	}$&$	-2.67^{+	0.21	}_{-	0.2	}$&$	58^{+	18	}_{-	19	}$&	2504(2457.1)&	7.5	\\
45	&	50	&$	-0.4^{+	0.5	}_{-	0.8	}$&$	-1.3^{+	0.8	}_{-0.5	}$&$	12.2^{+	3.4	}_{-	4	}$&$	 
49^{+	7	}_{-	6	}$&$	-2.47^{+	0.16	}_{-

0.17	}$&	3520.4(3485.1)&$	-0.3^{+	0.28	}_{-	0.29	}$&$	-1.54^{+	0.29	}_{-	0.28	 
}$&$	12.9^{+	3.2	}_{-	4	}$&$	44^{+	5	}_{-

6	}$&$	-2.53^{+	0.19	}_{-	0.18	}$&$	49^{+	17	}_{-	18	}$&	3512.7(3465.8)&	7.7	\\
50	&	70	&$	-0.2^{+	0.5	}_{-	0.5	}$&$	-1.64^{+	0.11	}_{-0.27	}$&$	11.2^{+	1.5	}_{-	 
1.8	}$&$	51^{+	8	}_{-	9	}$&$	-2.66^{+	

0.28	}_{-	0.3	}$&	4977.5(4942.3)&$	-0.1^{+	0.5	}_{-	0.5	}$&$	-1.66^{+	0.21	}_{-	0.19	 
}$&$	10.9^{+	1.6	}_{-	2.4	}$&$	43^{+	6	
}_{-	7	}$&$	-2.64^{+	0.25	}_{-	0.25	}$&$	42^{+	16	}_{-	17	}$&	4968.1(4921.2)&	 
9.4	\\

\hline
\end{tabular}
\label{fit_data}
\end{sidewaystable}

\begin{table*}
\centering
\renewcommand\tabcolsep{1.2pt}
\renewcommand\arraystretch{1.0}
\caption{The calculated flux and derived parameters of GRB 211211A prompt emission.}
\begin{tabular}{ccccccc} 
\hline\hline
$t_{1}$ & $t_{2}$ & $F_{\rm BB}$ & $F_{\rm obs}$ & $F_{\rm BB}/F_{\rm obs}$& $\Gamma_{\rm ph}$ & $R_{\rm ph}$
\\
$\rm (s)$ & $\rm (s)$ & $\rm (10^{-6}~erg~cm^{-2}~s^{-1}) $ & $\rm (10^{-6}~erg~cm^{-2}~s^{-1})$& $\quad $ &
$\quad$&$(10^{10}~\rm cm)$ \\
\hline
0.5	&	1.5	&$	0.03 	^{+	0.07 	}_{	-0.02 	}$&$	0.48 	^{+	0.06 	}_{	-0.06 	}$&$	0.06 	^{+	
0.15 	}_{-	0.05 	}$&$	86.93 	^{+	31.69 	}_{-

20.19 	}$&$	0.61 	^{+	0.66 	}_{-	0.42 	}$	\\
1.5	&	2.5	&$	0.32 	^{+	0.62 	}_{	-0.28 	}$&$	13.30 	^{+	0.30 	}_{	-0.20 	}$&$	0.02 	^{+	 
0.05 	}_{-	0.02 	}$&$	176.23 	^{+	49.35 	}_{-

31.14 	}$&$	2.02 	^{+	1.69 	}_{-	1.07 	}$	\\
2.5	&	3.5	&$	8.16 	^{+	1.54 	}_{	-1.66 	}$&$	54.40 	^{+	0.50 	}_{	-0.40 	}$&$	0.15 	^{+	 
0.03 	}_{-	0.03 	}$&$	268.67 	^{+	15.01 	}_{-

15.22 	}$&$	2.33 	^{+	0.39 	}_{-	0.40 	}$	\\
3.5	&	4.5	&$	0.91 	^{+	3.69 	}_{	-0.85 	}$&$	39.70 	^{+	0.30 	}_{	-0.70 	}$&$	0.02 	^{+	 
0.09 	}_{-	0.02 	}$&$	252.02 	^{+	137.33 	}_{-

58.45 	}$&$	2.06 	^{+	3.36 	}_{-	1.43 	}$	\\
4.5	&	5.5	&$	0.28 	^{+	0.65 	}_{	-0.25 	}$&$	13.90 	^{+	0.30 	}_{	-0.10 	}$&$	0.02 	^{+	 
0.05 	}_{-	0.02 	}$&$	192.10 	^{+	65.90 	}_{-

40.46 	}$&$	1.63 	^{+	1.67 	}_{-	1.03 	}$	\\
5.5	&	6.5	&$	2.98 	^{+	1.82 	}_{	-1.99 	}$&$	63.00 	^{+	1.00 	}_{	-0.30 	}$&$	0.05 	^{+	 
0.03 	}_{-	0.03 	}$&$	273.76 	^{+	48.23 	}_{-

49.10 	}$&$	2.55 	^{+	1.35 	}_{-	1.37 	}$	\\
6.5	&	7.5	&$	0.70 	^{+	2.20 	}_{	-0.66 	}$&$	81.50 	^{+	0.50 	}_{	-0.50 	}$&$	0.01 	^{+	 
0.03 	}_{-	0.01 	}$&$	311.73 	^{+	154.09 	}_{-

100.38 	}$&$	2.23 	^{+	3.31 	}_{-	2.16 	}$	\\
7.5	&	8.5	&$	4.80 	^{+	2.00 	}_{	-2.10 	}$&$	62.20 	^{+	0.50 	}_{	-0.40 	}$&$	0.08 	^{+	 
0.03 	}_{-	0.03 	}$&$	267.11 	^{+	27.37 	}_{-

27.73 	}$&$	2.71 	^{+	0.83 	}_{-	0.84 	}$	\\
8.5	&	9.5	&$	0.15 	^{+	0.88 	}_{	-0.14 	}$&$	15.10 	^{+	0.30 	}_{	-0.20 	}$&$	0.01 	^{+	 
0.06 	}_{-	0.01 	}$&$	221.40 	^{+	174.20 	}_{-

73.68 	}$&$	1.15 	^{+	2.72 	}_{-	1.15 	}$	\\
9.5	&	10.5	&$	0.08 	^{+	0.35 	}_{	-0.07 	}$&$	12.70 	^{+	0.20 	}_{	-0.20 	}$&$	0.01 	 
^{+	0.03 	}_{-	0.01 	}$&$	204.54 	^{+	117.81 	

}_{-	54.60 	}$&$	1.23 	^{+	2.13 	}_{-	0.99 	}$	\\
10.5	&	11.5	&$	0.28 	^{+	0.96 	}_{	-0.27 	}$&$	9.24 	^{+	0.19 	}_{	-0.17 	}$&$	0.03 	 
^{+	0.10 	}_{-	0.03 	}$&$	192.26 	^{+	
97.56

}_{-	58.06 	}$&$	1.08 	^{+	1.64 	}_{-	0.98 	}$	\\
11.5	&	12.5	&$	0.04 	^{+	0.12 	}_{	-0.03 	}$&$	3.85 	^{+	0.11 	}_{	-0.10 	}$&$	0.01 	 
^{+	0.03 	}_{-	0.01 	}$&$	148.01 	^{+	
66.21

}_{-	37.63 	}$&$	0.98 	^{+	1.32 	}_{-	0.75 	}$	\\
12.5	&	13.5	&$	0.03 	^{+	0.07 	}_{	-0.02 	}$&$	2.03 	^{+	0.08 	}_{	-0.08 	}$&$	0.01 	 
^{+	0.03 	}_{-	0.01 	}$&$	125.75 	^{+	
49.89

}_{-	28.54 	}$&$	0.85 	^{+	1.01 	}_{-	0.58 	}$	\\
13.5	&	15	&$	0.02 	^{+	0.06 	}_{	-0.02 	}$&$	2.20 	^{+	0.08 	}_{	-0.06 	}$&$	0.01 	 
^{+	0.03 	}_{-	0.01 	}$&$	127.76 	^{+	52.29 	

}_{-	30.06 	}$&$	0.87 	^{+	1.07 	}_{-	0.62 	}$	\\
15	&	17	&$	0.03 	^{+	0.09 	}_{	-0.03 	}$&$	3.90 	^{+	0.10 	}_{	-0.04 	}$&$	0.01 	^{+	 
0.02 	}_{-	0.01 	}$&$	145.94 	^{+	61.89 	}_{-

36.96 	}$&$	1.04 	^{+	1.32 	}_{-	0.79 	}$	\\
17	&	19	&$	0.10 	^{+	0.47 	}_{	-0.09 	}$&$	10.90 	^{+	0.10 	}_{	-0.20 	}$&$	0.01 	^{+	 
0.04 	}_{-	0.01 	}$&$	197.65 	^{+	133.22 	}_{-

63.08 	}$&$	1.17 	^{+	2.37 	}_{-	1.12 	}$	\\
19	&	21	&$	0.14 	^{+	0.30 	}_{	-0.12 	}$&$	9.57 	^{+	0.13 	}_{	-0.14 	}$&$	0.01 	^{+	 
0.03 	}_{-	0.01 	}$&$	173.16 	^{+	57.02 	}_{-

37.15 	}$&$	1.53 	^{+	1.51 	}_{-	0.98 	}$	\\
21	&	23	&$	0.25 	^{+	0.42 	}_{	-0.22 	}$&$	9.22 	^{+	0.12 	}_{	-0.13 	}$&$	0.03 	^{+	 
0.05 	}_{-	0.02 	}$&$	169.90 	^{+	46.13 	}_{-

34.54 	}$&$	1.56 	^{+	1.27 	}_{-	0.95 	}$	\\
23	&	25	&$	0.13 	^{+	0.35 	}_{	-0.12 	}$&$	10.70 	^{+	0.13 	}_{	-0.17 	}$&$	0.01 	^{+	 
0.03 	}_{-	0.01 	}$&$	182.36 	^{+	69.38 	}_{-

38.22 	}$&$	1.46 	^{+	1.67 	}_{-	0.92 	}$	\\
25	&	27	&$	0.07 	^{+	0.33 	}_{	-0.06 	}$&$	6.40 	^{+	0.12 	}_{	-0.12 	}$&$	0.01 	^{+	 
0.05 	}_{-	0.01 	}$&$	175.75 	^{+	121.45 	}_{-

59.16 	}$&$	0.98 	^{+	2.03 	}_{-	0.99 	}$	\\
27	&	29	&$	0.38 	^{+	0.26 	}_{	-0.31 	}$&$	3.36 	^{+	0.08 	}_{	-0.08 	}$&$	0.11 	^{+	 
0.08 	}_{-	0.09 	}$&$	125.82 	^{+	16.68 	}_{-

17.90 	}$&$	1.40 	^{+	0.56 	}_{-	0.60 	}$	\\
29	&	31	&$	0.02 	^{+	0.06 	}_{	-0.02 	}$&$	2.81 	^{+	0.07 	}_{	-0.06 	}$&$	0.01 	^{+	 
0.02 	}_{-	0.01 	}$&$	138.25 	^{+	51.92 	}_{-

29.94 	}$&$	0.88 	^{+	0.99 	}_{-	0.57 	}$	\\
31	&	33	&$	0.17 	^{+	0.17 	}_{	-0.15 	}$&$	2.43 	^{+	0.06 	}_{	-0.07 	}$&$	0.07 	^{+	 
0.07 	}_{-	0.06 	}$&$	122.80 	^{+	25.81 	}_{-

24.50 	}$&$	1.09 	^{+	0.69 	}_{-	0.65 	}$	\\
33	&	35	&$	0.04 	^{+	0.12 	}_{	-0.03 	}$&$	2.28 	^{+	0.07 	}_{	-0.08 	}$&$	0.02 	^{+	 
0.05 	}_{-	0.02 	}$&$	128.03 	^{+	56.02 	}_{-

30.89 	}$&$	0.90 	^{+	1.18 	}_{-	0.65 	}$	\\
35	&	37	&$	0.08 	^{+	0.19 	}_{	-0.07 	}$&$	3.36 	^{+	0.10 	}_{	-0.08 	}$&$	0.02 	^{+	 
0.06 	}_{-	0.02 	}$&$	143.44 	^{+	55.25 	}_{-

37.13 	}$&$	0.94 	^{+	1.09 	}_{-	0.73 	}$	\\
37	&	39	&$	0.06 	^{+	0.17 	}_{	-0.05 	}$&$	2.46 	^{+	0.08 	}_{	-0.08 	}$&$	0.02 	^{+	 
0.07 	}_{-	0.02 	}$&$	128.31 	^{+	54.07 	}_{-

30.66 	}$&$	0.97 	^{+	1.22 	}_{-	0.69 	}$	\\
39	&	41	&$	0.05 	^{+	0.16 	}_{	-0.04 	}$&$	1.95 	^{+	0.06 	}_{	-0.07 	}$&$	0.02 	^{+	 
0.08 	}_{-	0.02 	}$&$	123.11 	^{+	60.78 	}_{-

33.25 	}$&$	0.87 	^{+	1.28 	}_{-	0.70 	}$	\\
41	&	43	&$	0.48 	^{+	0.11 	}_{	-0.14 	}$&$	1.74 	^{+	0.07 	}_{	-0.07 	}$&$	0.28 	^{+	 
0.06 	}_{-	0.08 	}$&$	120.34 	^{+	9.17 	}_{-

9.59 	}$&$	0.83 	^{+	0.19 	}_{-	0.20 	}$	\\
43	&	45	&$	0.08 	^{+	0.10 	}_{	-0.07 	}$&$	1.05 	^{+	0.05 	}_{	-0.04 	}$&$	0.08 	^{+	 
0.09 	}_{-	0.06 	}$&$	104.86 	^{+	22.62 	}_{-

19.58 	}$&$	0.76 	^{+	0.49 	}_{-	0.42 	}$	\\
45	&	50	&$	0.04 	^{+	0.08 	}_{	-0.03 	}$&$	1.16 	^{+	0.03 	}_{	-0.04 	}$&$	0.03 	^{+	 
0.07 	}_{-	0.03 	}$&$	108.06 	^{+	32.72 	}_{-

22.00 	}$&$	0.76 	^{+	0.69 	}_{-	0.47 	}$	\\
50	&	70	&$	0.02 	^{+	0.04 	}_{	-0.01 	}$&$	0.52 	^{+	0.02 	}_{	-0.02 	}$&$	0.03 	^{+	 
0.08 	}_{-	0.02 	}$&$	91.12 	^{+	33.68 	}_{-

18.95 	}$&$	0.57 	^{+	0.63 	}_{-	0.36 	}$	\\

\hline
\end{tabular}
\label{fit_data}
\end{table*}

\begin{table*}
\centering
\renewcommand\tabcolsep{1.2pt}
\renewcommand\arraystretch{1.0}
\caption{Derived parameters of GRB 211211A in a hybrid jet model.}
\begin{tabular}{cccccccc} 
\hline\hline
$t_{1}$&$t_{2}$&$1+\sigma_{0}$&$\eta$&$R_{\rm ph}$&$\Gamma_{\rm ph}$&$1+\sigma_{\rm ph}$\\
$(\rm s)$&$(\rm s)$&$\quad $&$\quad $&($10^{10}$ cm)&$\quad $&$\quad $\\
\hline
0.5	&	1.5	&	11.57 	&	26.46 	&	0.51 	&	71.17 	&	4.31 	\\
1.5	&	2.5	&	21.71 	&	38.16 	&	1.54 	&	131.05 	&	6.34 	\\
2.5	&	3.5	&	4.18 	&	97.49 	&	2.59 	&	291.33 	&	1.40 	\\
3.5	&	4.5	&	28.45 	&	57.89 	&	1.41 	&	167.94 	&	9.83 	\\
4.5	&	5.5	&	31.82 	&	42.32 	&	1.09 	&	125.05 	&	10.79 	\\
5.5	&	6.5	&	12.63 	&	73.34 	&	2.17 	&	227.15 	&	4.09 	\\
6.5	&	7.5	&	82.86 	&	57.57 	&	1.15 	&	156.42 	&	30.56 	\\
7.5	&	8.5	&	7.31 	&	79.44 	&	2.68 	&	256.82 	&	2.27 	\\
8.5	&	9.5	&	87.35 	&	45.45 	&	0.56 	&	105.38 	&	37.76 	\\
9.5	&	10.5	&	122.25 	&	36.58 	&	0.56 	&	91.13 	&	49.18 	\\
10.5	&	11.5	&	27.43 	&	51.29 	&	0.71 	&	123.36 	&	11.43 	\\
11.5	&	12.5	&	76.10 	&	28.70 	&	0.51 	&	75.17 	&	29.12 	\\
12.5	&	13.5	&	59.24 	&	25.84 	&	0.47 	&	68.10 	&	22.53 	\\
13.5	&	15	&	73.64 	&	24.71 	&	0.46 	&	65.66 	&	27.77 	\\
15	&	17	&	89.47 	&	26.48 	&	0.53 	&	71.80 	&	33.07 	\\
17	&	19	&	91.10 	&	38.27 	&	0.58 	&	94.59 	&	36.94 	\\
19	&	21	&	42.91 	&	34.85 	&	0.95 	&	105.14 	&	14.25 	\\
21	&	23	&	22.60 	&	39.56 	&	1.15 	&	121.69 	&	7.36 	\\
23	&	25	&	54.28 	&	35.86 	&	0.85 	&	103.18 	&	18.91 	\\
25	&	27	&	80.90 	&	35.89 	&	0.49 	&	85.97 	&	33.85 	\\
27	&	29	&	4.65 	&	40.46 	&	1.57 	&	137.02 	&	1.38 	\\
29	&	31	&	94.71 	&	25.78 	&	0.43 	&	66.13 	&	37.00 	\\
31	&	33	&	8.88 	&	36.59 	&	1.01 	&	110.63 	&	2.94 	\\
33	&	35	&	42.08 	&	28.19 	&	0.55 	&	75.94 	&	15.66 	\\
35	&	37	&	32.97 	&	34.42 	&	0.61 	&	89.68 	&	12.68 	\\
37	&	39	&	29.91 	&	30.00 	&	0.65 	&	83.59 	&	10.76 	\\
39	&	41	&	30.64 	&	29.34 	&	0.57 	&	79.06 	&	11.40 	\\
41	&	43	&	2.63 	&	54.13 	&	1.00 	&	141.83 	&	$\cdots$	\\
43	&	45	&	9.28 	&	33.40 	&	0.67 	&	91.03 	&	3.41 	\\
45	&	50	&	21.26 	&	28.19 	&	0.55 	&	76.05 	&	7.90 	\\
50	&	70	&	24.26 	&	24.05 	&	0.39 	&	61.14 	&	9.56 	\\

\hline
\end{tabular}
\end{table*}
\label{fit_data}

\clearpage




\begin{thebibliography}{99}
\bibitem[Abdo et al.(2009)]{2009ApJ...706L.138A} Abdo, A.~A., Ackermann, M., Ajello, M., et al.\ 2009, \apjl, 
    706, L138. doi:10.1088/0004-637X/706/1/L138
\bibitem[Axelsson et al.(2012)]{2012ApJ...757L..31A} Axelsson, M., Baldini, L., Barbiellini, G., et al.\ 
    2012, \apjl, 757, L31. doi:10.1088/2041-8205/757/2/L31
\bibitem[Band et al.(1993)]{1993ApJ...413..281B} Band, D., Matteson, J., Ford, L., et al.\ 1993,
    \apj, 413, 281. doi:10.1086/172995
\bibitem[Berger(2014)]{Berger2014} Berger, E.\ 2014, \araa, 52, 43.
\bibitem[Blandford \& Znajek(1977)]{1977MNRAS.179..433B} Blandford, R.~D. \& Znajek, R.~L.\ 1977, \mnras, 
    179, 433. doi:10.1093/mnras/179.3.433
\bibitem[Bucciantini et al.(2012)]{Bucciantini2012} Bucciantini, N., Metzger, B.~D., Thompson,
    T.~A., et al.\ 2012, \mnras, 419, 1537. doi:10.1111/j.1365-2966.2011.19810.x
\bibitem[D'Ai et al.(2021)]{2021GCN.31202....1D} D'Ai, A., Ambrosi, E., D'Elia, V., et al.\ 2021,
    GRB Coordinates Network, Circular Service, No. 31202, 1
\bibitem[Dai \& Lu(1998a)]{Dai1998a} Dai, Z.~G. \& Lu, T.\ 1998a, \mnras, 298, 87.
    doi:10.1046/j.1365-8711.1998.01681.x
\bibitem[Dai \& Lu(1998b)]{Dai1998b} Dai, Z.~G. \& Lu, T.\ 1998b, \aap, 333, L87
\bibitem[Eichler et al.(1989)]{Eichler1989} Eichler, D., Livio, M., Piran, T., et al.\
    1989, \nat, 340, 126. doi:10.1038/340126a0
\bibitem[Fenimore et al.(1993)]{1993A&AS...97...59F} Fenimore, E.~E., Epstein, R.~I., \& Ho, C.\
    1993, \aaps, 97, 59
\bibitem[Gao \& Zhang (2015)]{2015ApJ...801...2} Gao, H. \& Zhang B et al.\ 2015, \apj,
    801, 5. doi:0.1088/0004-637X/801/2/103
\bibitem[Gao et al.(2022)]{Gao2022} Gao, H., Lei, W.-H., \& Zhu, Z.-P.\ 2022, \apjl, 934, L12. 
    doi:10.3847/2041-8213/ac80c7
\bibitem[Gompertz et al.(2022)]{Gompertz2022} Gompertz, B.~P., Ravasio, M.~E., Nicholl, M.,
    et al.\ 2022, Nature Astronomy. doi:10.1038/s41550-022-01819-4
\bibitem[Goodman(1986)]{1986ApJ...308L..47G} Goodman, J.\ 1986, \apjl, 308, L47. doi:10.1086/184741
\bibitem[Guiriec et al.(2011)]{2011ApJ...727L..33G} Guiriec, S., Connaughton, V., Briggs, M.~S., et al.\ 
    2011, \apjl, 727, L33. doi:10.1088/2041-8205/727/2/L33
\bibitem[Guiriec et al.(2013)]{2013ApJ...770...32G} Guiriec, S., Daigne, F., Hasco{\"e}t, R., et al.\ 2013, 
    \apj, 770, 32. doi:10.1088/0004-637X/770/1/32
\bibitem[Gu et al.(2006)]{2006ApJ...643L..87G} Gu, W.-M., Liu, T., \& Lu, J.-F.\ 2006, \apjl, 643, L87. 
    doi:10.1086/505140
\bibitem[Hou et al.(2018)]{2018ApJ...866...13H} Hou, S.-J., Zhang, B.-B., Meng, Y.-Z., et al.\
    2018, \apj, 866, 13. doi:10.3847/1538-4357/aadc07
\bibitem[Kumar \& Zhang(2015)]{2015PhR...561....1K} Kumar, P. \& Zhang, B.\ 2015, \physrep, 561, 1.
    doi:10.1016/j.physrep.2014.09.008
\bibitem[Kobayashi \& Zhang(2007)]{2007ApJ...655..973K} Kobayashi, S. \& Zhang, B.\ 2007, \apj, 655, 973. 
    doi:10.1086/510203
\bibitem[Lee et al.(2000)]{2000PhR...325...83L} Lee, H.~K., Wijers, R.~A.~M.~J., \& Brown, G.~E.\ 2000, 
    \physrep, 325, 83. doi:10.1016/S0370-1573(99)00084-8
\bibitem[Lei et al.(2009)]{2009ApJ...700.1970L} Lei, W.~H., Wang, D.~X., Zhang, L., et al.\ 2009, \apj, 700, 
    1970. doi:10.1088/0004-637X/700/2/1970
\bibitem[Lei et al.(2013)]{2013ApJ...765..125L} Lei, W.-H., Zhang, B., \& Liang, E.-W.\ 2013, \apj, 765, 125. 
    doi:10.1088/0004-637X/765/2/125
\bibitem[Lei et al.(2017)]{2017ApJ...849...47L} Lei, W.-H., Zhang, B., Wu, X.-F., et al.\ 2017, \apj, 849, 
    47. doi:10.3847/1538-4357/aa9074
\bibitem[Li \& Paczy{\'n}ski(2000)]{2000ApJ...534L.197L} Li, L.-X. \& Paczy{\'n}ski, B.\ 2000,
    \apjl, 534, L197. doi:10.1086/312678
\bibitem[Li(2019)]{Li2019} Li, L.\ 2019, \apjs, 242, 16.
\bibitem[Liang et al.(2010)]{2010ApJ...725.2209L} Liang, E.-W., Yi, S.-X., Zhang, J., et al.\ 2010,
    \apj, 725, 2209. doi:10.1088/0004-637X/725/2/2209
\bibitem[Lithwick \& Sari(2001)]{2001ApJ...555..540L} Lithwick, Y. \& Sari, R.\ 2001, \apj, 555, 540. 
    doi:10.1086/321455
\bibitem[L{\"u} et al.(2017)]{2017ApJ...849...71L} L{\"u}, H.-J., L{\"u}, J., Zhong, S.-Q., et al.\ 2017, 
    \apj, 849, 71. doi:10.3847/1538-4357/aa8f99
\bibitem[L{\"u} \& Zhang(2014)]{Lv2014} L{\"u}, H.-J. \& Zhang, B.\ 2014, \apj, 785,
    74. doi:10.1088/0004-637X/785/1/74
\bibitem[L{\"u} et al.(2015)]{Lv2015} L{\"u}, H.-J., Zhang, B., Lei, W.-H., et al.\
    2015, \apj, 805, 89. doi:10.1088/0004-637X/805/2/89
\bibitem[L{\"u} et al.(2022)]{2022ApJ...931L..23L} L{\"u}, H.-J., Yuan, H.-Y., Yi, T.-F., et al.\
    2022, \apjl, 931, L23. doi:10.3847/2041-8213/ac6e3a
\bibitem[L{\"u} et al.(2012)]{2012ApJ...751...49L} L{\"u}, J., Zou, Y.-C., Lei, W.-H., et al.\
    2012, \apj, 751, 49. doi:10.1088/0004-637X/751/1/49
\bibitem[Kaneko et al.(2006)]{2006ApJS..166..298K} Kaneko, Y., Preece, R.~D., Briggs, M.~S., et
    al.\ 2006, \apjs, 166, 298.
\bibitem[Mangan et al.(2021)]{2021GCN.31210....1M} Mangan, J., Dunwoody, R., Meegan, C., et al.\
    2021, GRB Coordinates Network, Circular Service, No. 31210, 1
\bibitem[Meegan et al.(2009)]{2009ApJ...702..791M} Meegan, C., Lichti, G., Bhat, P.~N., et al.\ 2009, \apj, 
    702, 791. doi:10.1088/0004-637X/702/1/791
\bibitem[Metzger et al.(2011)]{Metzger2011} Metzger, B.~D., Giannios, D., Thompson, T.~A.,
    et al.\ 2011, \mnras, 413, 2031. doi:10.1111/j.1365-2966.2011.18280.x
\bibitem[M{\'e}sz{\'a}ros \& Rees(1997)]{1997ApJ...476..232M} M{\'e}sz{\'a}ros, P. \& Rees, M.~J.\ 1997, 
    \apj, 476, 232. doi:10.1086/303625
\bibitem[M{\'e}sz{\'a}ros(2002)]{2002ARA&A..40..137M} M{\'e}sz{\'a}ros, P.\ 2002, \araa, 40, 137. 
    doi:10.1146/annurev.astro.40.060401.093821
\bibitem[Narayan et al.(2001)]{Narayan2001} Narayan, R., Piran, T., \& Kumar, P.\
    2001, \apj, 557, 949. doi:10.1086/322267

\bibitem[Neath \& Cavanaugh(2012)]{Neath2012} Neath, A.~A. \& Cavanaugh, J.~E.\ 2012, WIREs
Comput. Stat., 4, 199. doi: 10.1002/wics.199

\bibitem[Paczynski(1986)]{1986ApJ...308L..43P} Paczynski, B.\ 1986, \apjl, 308, L43.
    doi:10.1086/184740
\bibitem[Pe'er et al.(2007)]{2007ApJ...664L...1P} Pe'er, A., Ryde, F., Wijers, R.~A.~M.~J., et al.\ 2007, 
    \apjl, 664, L1. doi:10.1086/520534
\bibitem[Popham et al.(1999)]{1999ApJ...518..356P} Popham, R., Woosley, S.~E., \& Fryer, C.\ 1999,
    \apj, 518, 356. doi:10.1086/307259
\bibitem[Ravasio et al.(2018)]{2018A&A...613A..16R} Ravasio, M.~E., Oganesyan, G., Ghirlanda, G.,
    et al.\ 2018, \aap, 613, A16.
\bibitem[Rastinejad et al.(2022)]{Rastinejad2022} Rastinejad, J.~C., Gompertz, B.~P., Levan,
    A.~J., et al.\ 2022, \nat, 612, 223. doi:10.1038/s41586-022-05390-w
\bibitem[Rees \& Meszaros(1994)]{1994ApJ...430L..93R} Rees, M.~J. \& Meszaros, P.\ 1994, \apjl, 430, L93. 
    doi:10.1086/187446
\bibitem[Ryde(2005)]{2005ApJ...625L..95R} Ryde, F.\ 2005, \apjl, 625, L95. doi:10.1086/431239
\bibitem[Ryde et al.(2010)]{2010ApJ...709L.172R} Ryde, F., Axelsson, M., Zhang, B.~B., et al.\
    2010, \apjl, 709, L172. doi:10.1088/2041-8205/709/2/L172
\bibitem[Sari et al.(1998)]{1998ApJ...497L..17S} Sari, R., Piran, T., \& Narayan, R.\ 1998, \apjl, 497, L17. 
    doi:10.1086/311269
\bibitem[Sari \& Piran(1999)]{1999A&AS..138..537S} Sari, R. \& Piran, T.\ 1999, \aaps, 138, 537. 
    doi:10.1051/aas:1999342
\bibitem[Thompson(1994)]{Thompson1994} Thompson, C.\ 1994, \mnras, 270, 480.
    doi:10.1093/mnras/270.3.480
\bibitem[Usov(1992)]{Usov1992} Usov, V.~V.\ 1992, \nat, 357, 472. doi:10.1038/357472a0
\bibitem[Vianello et al.(2017)]{2017ifs..confE.130V} Vianello, G., Lauer, R.~J., Burgess, J.~M., et al.\ 
    2017, Proceedings of the 7th International Fermi Symposium,
    130
\bibitem[Wang et al.(2017)]{2017ApJ...836...81W} Wang, Y.-Z., Wang, H., Zhang, S., et al.\ 2017, \apj, 836, 
    81. doi:10.3847/1538-4357/aa56c6
\bibitem[Xiao et al.(2022)]{xiao2022} Xiao, S., Zhang, Y,-Q., Zhu, Z,-P., et al.2022, arXiv:2205.02186
\bibitem[Yang et al.(2022)]{yang2022} Yang, J., Ai, S., Zhang, B.-B., et al.\ 2022,
    \nat, 612, 232. doi:10.1038/s41586-022-05403-8
\bibitem[Zhang(2018)]{2018pgrb.book.....Z} Zhang, B.\ 2018, The Physics of Gamma-Ray Bursts by Bing Zhang. 
    ISBN: 978-1-139-22653-0. Cambridge Univeristy Press, 2018.
    doi:10.1017/9781139226530
\bibitem[Zhang \& Pe'er(2009)]{2009ApJ...700L..65Z} Zhang, B. \& Pe'er, A.\ 2009, \apjl, 700, L65.
    doi:10.1088/0004-637X/700/2/L65
\bibitem[Zhang \& M{\'e}sz{\'a}ros(2001)]{Zhang2001} Zhang, B. \& M{\'e}sz{\'a}ros, P.\
    2001, \apjl, 552, L35. doi:10.1086/320255
\bibitem[Zhang \& M{\'e}sz{\'a}ros(2004)]{2004IJMPA..19.2385Z} Zhang, B. \& M{\'e}sz{\'a}ros, P.\
    2004, International Journal of Modern Physics A, 19, 2385. doi:10.1142/S0217751X0401746X
\bibitem[Zhang \& Yan(2011)]{2011ApJ...726...90Z} Zhang, B. \& Yan, H.\ 2011, \apj, 726, 90.
    doi:10.1088/0004-637X/726/2/90
\bibitem[Zhang et al.(2016)]{Zhang2016} Zhang, B., L{\"u}, H.-J., \& Liang, E.-W.\ 2016,
    \ssr, 202, 3. doi:10.1007/s11214-016-0305-9
\bibitem[Zhang et al.(2011)]{2011ApJ...730..141Z} Zhang, B.-B., Zhang, B., Liang, E.-W., et al.\
    2011, \apj, 730, 141. doi:10.1088/0004-637X/730/2/141
\bibitem[Zhang et al.(2018)]{2018NatAs...2...69Z} Zhang, B.-B., Zhang, B., Castro-Tirado, A.~J., et
    al.\ 2018, Nature Astronomy, 2, 69. doi:10.1038/s41550-017-0309-8
\bibitem[Zhang et al.(2021)]{2021GCN.31236....1Z} Zhang, Y.~Q., Xiong, S.~L., Li, X.~B., et al.\
    2021, GRB Coordinates Network, Circular Service, No. 31236, 1
\end{thebibliography}
\end{document}